\documentclass[pre,aps,reprint,showpacs,preprintnumbers,superscriptaddress]{revtex4-1}
\usepackage{graphicx}
\usepackage{dcolumn}
\usepackage{bm}
\usepackage{amssymb}
\usepackage{pifont}
\usepackage{amsmath}
\usepackage{times}
\usepackage[caption=false]{subfig}
\usepackage[dvipsnames]{xcolor}
\usepackage{multirow}

\begin{document}
\title{Dynamics of a large system of spiking neurons with synaptic delay}
\author{Federico Devalle}
\affiliation{Neuronal Dynamics Group, Department of Information and Communication Technologies, Universitat Pompeu Fabra, 08018 Barcelona, Spain}
\affiliation{Department of Physics, Lancaster University, Lancaster LA1 4YB, United Kingdom}
\author{Ernest Montbri\'o}
\affiliation{Neuronal Dynamics Group, Department of Information and Communication Technologies, Universitat Pompeu Fabra, 08018 Barcelona, Spain}
\author{Diego Paz\'o}
\affiliation{Inst\'ituto de Fisica de Cantabria (IFCA), CSIC-Universidad de Cantabria, 39005 Santander, Spain}
\date{\today}

\begin{abstract}
We analyze a large system of heterogeneous
quadratic integrate-and-fire (QIF)
neurons with time delayed, all-to-all synaptic
coupling. The model is exactly reduced to a system of firing
rate equations that is exploited to investigate the existence,
stability and bifurcations of fully synchronous, partially synchronous,
and incoherent states.
In conjunction with this analysis we perform
extensive numerical simulations of the original network of QIF neurons,
and determine the relation between the
macroscopic and microscopic states for partially synchronous states.  
The results are summarized in two phase diagrams, for homogeneous and
heterogeneous populations, which are obtained analytically to a large extent.
For excitatory coupling, the phase diagram is remarkably similar to that of the
Kuramoto model with time delays, although here the stability boundaries 
extend to regions in parameter space where the neurons are not
self-sustained oscillators. In contrast, the structure of the boundaries
for inhibitory coupling is different, and already for homogeneous networks
unveils the presence of various partially synchronized states
not present in the Kuramoto model:
Collective chaos, quasiperiodic partial synchronization (QPS), and a novel
state which we call modulated-QPS (M-QPS).
In the presence of heterogeneity partially synchronized states reminiscent to 
collective chaos, QPS and M-QPS persist. In addition, the presence of 
heterogeneity greatly amplifies the differences between the incoherence 
stability boundaries of excitation and inhibition. Finally, 
we compare our results with those of a traditional (Wilson Cowan-type)
firing rate model 
with time delays. The oscillatory instabilities of the
traditional firing rate model qualitatively agree with our results
only for the case of inhibitory coupling with strong heterogeneity.
\end{abstract}
\maketitle

\section{Introduction}

Since the seminal work of Hodgkin and Huxley~\cite{HH52},
spiking neuron models have been the standard mathematical tool
to investigate the collective dynamics of neuronal networks.
These models account for the basic properties of neurons
---sub-threshold voltage dynamics, spiking,
and discontinuous synaptic interactions--- and hence
networks of spiking neurons are considered to be biologically realistic. 
Yet, network models of spiking neurons are generally not amenable to
analysis and hence mostly constitute a computational tool.

Alternatively, researchers use simplified models which
describe some measure of the mean activity in a
population of cells, customarily taken as the firing
rate~\cite{WC72}. Such mean field models (that here we call `traditional
firing rate models' or simply `firing rate models') faithfully capture the
main types of qualitative dynamical states observed in large populations
of asynchronously spiking neurons, and
can be mathematically analyzed using
standard techniques for differential equations,
see e.g.~\cite{DA01,GK02,ET10,ACN16}.
Despite their popularity, 
traditional firing rate models have two major limitations which
strongly limit their range of applicability in neuroscience.
First, these models are not accurate in
describing the dynamics of collective
states where a significant fraction of the
neurons fires spikes in synchrony.
Second, firing rate models do not generally represent proper
mathematical reductions of the original network but rather
are heuristic. As such there is in general
no precise relationship between the
parameters in the traditional firing rate model and those in the full
network of spiking
neurons, and thus there is no clear link between the
macroscopic states of the network with
the microscopic dynamics of the constituent neurons. 

An important example of the application of traditional firing
rate models occurs in the analysis of neuronal networks with time delays.
It is well-known that synaptic and dendritic
processing, as well as axonal propagation, produce unavoidable
time delays in the neuronal interactions which profoundly
shape the oscillatory dynamics of spiking neuron networks.
The study of large networks of spiking neurons
with time delays is convoluted, and in this context
the mathematical and numerical analysis of
firing rate descriptions have been particularly
productive, see e.g.~\cite{RBH05,BBH07,BH08,RM11,LB11,KFR17,
KEK18,SKS+18,HA06,BK08,VCM07,CL09,FF10,Tou12,WRO+12,Vel13,VF13,FT14,DGJ+15}.
Yet, how much of the
actual dynamics of a large network of spiking neurons with synaptic delays
can be captured using traditional firing rate descriptions?

In this paper, we investigate the collective dynamics of a large system of
heterogeneous quadratic integrate-and-fire (QIF) neurons with synaptic
delays. To perform the analysis we exploit a novel
low-dimensional firing rate model
that can be exactly derived from the population of QIF neurons~\cite{MPR15}.
Therefore we use a system of firing rate equations
(FRE) that, in contrast with traditional firing rate models,
faithfully reproduce all possible collective states of the network.
The mathematical analysis of the FRE
allows us to obtain exact formulas for the boundaries of stability of
asynchronous states in both homogeneous and heterogeneous networks.
In conjunction with this analysis, we conduct numerical simulations
of the corresponding network of QIF neurons in order to investigate
the microscopic states associated with the macroscopic dynamics of the FRE.
This combined analysis
reveals the presence of large regions of oscillatory states which
are unreachable using traditional firing rate models.
Some of these states are particularly interesting and we
investigate them in detail. They
already arise in populations of identical inhibitory neurons, in parameter
regions where both the fully synchronous and the asynchronous  
states are unstable. Hence, in these regimes, the system settles somewhere
in between full order and disorder,
at a state often called `partial synchrony'
\footnote{To be consistent with previous work
investigating the collective dynamics of
identical oscillators
---see the review paper~\cite{PR15i} and references therein---,
we adopt the term `partial synchronization' to refer to
states which are neither fully synchronous nor asynchronous
(excluding cluster states).
We note that the same term is used
to designate synchronous states in the Kuramoto model with heterogeneity.
Here we use the same terminology for both types of states (see Section V).}.
Such partially synchronous states in networks of identical
units are self-organized collective states in which the properties 
of the mean field cannot be
trivially inferred from the intrinsic dynamics of the units,
but are an emergent property of the network. Here we find three different
types of partially synchronous states (with periodic, quasiperiodic, and
chaotic mean field dynamics) and also investigate how these states change
as neurons are made heterogeneous.

Our work builds primarily on the results by two of the authors~\cite{PM16}
about the dynamics of networks of identical, self-oscillatory QIF neurons.
Here we extend the results in~\cite{PM16}
in several ways:
\begin{itemize}
\item The analysis is not restricted to self-oscillating QIF neurons,
but extends to networks of excitable QIF neurons.
\item We perform a detailed numerical exploration of the
partially synchronous states and their bifurcations,
supported by the systematic computation of the Lyapunov exponents.
This allows us to uncover a
transition to a novel state which we call modulated quasiperiodic
partial synchronization (M-QPS), as well as a `quasiperiodic route'
to collective chaos. Additionally we investigate
how partially synchronous states transform as neurons are made heterogeneous.
To the best of our knowledge, this problem has not been addressed in previous
work investigating partial synchronization
in different populations of identical oscillators 
~\cite{NK95,Vre96,VD03,MP06,OPT10,LOP+12,RP07,BSV+14,temirbayev12,temirbayev13,
PR15,CRP16,RP15}.
\item We obtain the phase diagram corresponding to populations of
heterogeneous QIF neurons. Heterogeneity magnifies the
difference between the dynamics of inhibitory and excitatory networks.
The phase diagram is finally compared
with that of a traditional firing rate model,
which we heuristically obtain from the
exact FRE obtained in~\cite{MPR15}.
The oscillatory instabilities of the
two firing rate models qualitatively agree only for the case of
inhibitory networks with strong heterogeneity.
\end{itemize}

The paper is organized as follows.
In Sec.~II we present the time delayed QIF network model under study.
In Sec.~III we introduce and discuss the low-dimensional FRE
derived, in the large system-size limit, from the QIF network.
In Sec.~IV we complement the theoretical analysis
of the FRE with numerical simulations of the 
partially synchronous states (QPS, M-QPS, collective chaos).
In Sec.~V we analyze the effect of heterogeneities in the system dynamics.
Finally, in Sec.~VI we discuss our results and compare them 
with those obtained using a traditional firing rate description.

\section{Model Description}
\label{sec:model}
We consider a network of $N\gg1$ all-to-all coupled QIF neurons.
The membrane potential of the neurons is governed by the following
quadratic differential equation \cite{Izh07}
\begin{equation}\label{qif}
\tau\dot{V}_{j}=V^2_{j}+I_{j} \qquad j=1,\ldots ,N
\end{equation}
where $\tau$ is a time constant.
Every time the membrane potential of a neuron reaches an upper
threshold $V_\text{th}\gg1$ it is said to fire.
Obviously, in addition to \eqref{qif}, one must define
a spike-resetting condition
\begin{equation}\label{eq=reset}
\text{If} \quad V_{j}>V_{\text{th}} \quad \text{then} \quad V_{\text{reset}} \leftarrow V_{j}.
\end{equation}
In our theoretical analysis we consider the limits $V_{\text{th}}=-V_{\text{reset}}\rightarrow \infty$,
which is faithfully reproduced in numerical simulations in the following way:
first, we consider $V_{\text{th}}=-V_{\text{reset}}=500$.
Then, after the firing, we set the neuron at $V_{\text{reset}}$
after an inactive period of $2 \tau / V_{\text{th}}$.
This is the approximate time that a neuron needs
to reach $+\infty$ from $V_{\text{th}}$
and return from $-\infty$ to $V_{\text{reset}}$
\footnote{For the numerical simulations of the
population of QIF neurons we use the Euler method
with time step $\delta t=10^{-5}$. For the integration of
the FREs Eqs.~(\ref{fre1}) we use a third-order Adams-Bashforth-Moulton predictor-corrector scheme with a timestep $\delta t=10^{-4}$~\cite{Press}. In all simulations shown, initial transients were discarded.}.

The input in Eq.~\eqref{qif} is determined by two distinct contributions:
\begin{equation}\label{eq=QIF_current}
I_{j}=\eta_{j}+Js\left(t\right).
\end{equation}
The first term represents the quenched heterogeneity,
which for neurons in the oscillatory regime ($\eta_j>0$),
determines the intrinsic interspike interval (ISI)
\begin{equation}
T_j=\pi\tau/\sqrt{\eta_{j}}.
\label{isi}
\end{equation}
The second term corresponds to the mean field coupling, where $J$ is the coupling strength and $s\left(t\right)$ is the mean synaptic activation.
We consider networks of spiking neurons with delayed,
mean-field coupling
\begin{equation}\label{eq=syn_mf}
s(t)=\dfrac{\tau}{N\tau_{s}}\sum_{j=1}^{N}\sum_{k}\int_{t-D-\tau_{s}}^{t-D}\delta\left(t^{\prime}-t_{j}^{k}\right)dt^{\prime}.
\end{equation}
where $t_{j}^{k}$ is the time of the $k\text{th}$ spike of neuron $j$,
and $\tau_{s}$ the synaptic time constant.
After adopting the thermodynamic limit, $N\to\infty$,
we take the limit $\tau_s\to0$, so that $s$ becomes
proportional to the instantaneous population-averaged firing
rate at time $t-D$:
$$\lim_{\tau_s \to 0} \lim_{N\to \infty} s(t)=\tau \,r(t-D)\equiv \tau \, r_D.$$
Finally, we assume a Lorentzian (Cauchy) distribution of the quenched heterogeneity
\begin{equation}
g\left(\eta\right)=\dfrac{\Delta /\pi}{\left(\eta-\bar{\eta}\right)^2+\Delta^2}.
\label{eq=lorentzian}
\end{equation}

\section{Low-dimensional description: Firing rate equations}

In the thermodynamic limit, the network of QIF neurons can be reduced
to a finite set of FRE~\cite{MPR15,PD16}. This is possible
assuming that the conditional neuron
densities $\rho(V|\eta,t)$ are Lorentzian for all $\eta$ values~\cite{MPR15},
which is mathematically equivalent as to invoke
the so-called Ott-Antonsen (OA) theory~\cite{OA08}.

Specifically, the original work by Ott \& Antonsen
applies to the Kuramoto model, and it
shows that the model admits an exact,
low-dimensional description in terms of the 
Kuramoto order parameter~\cite{OA08}. The
same theory holds for large populations of globally
pulse-coupled oscillators~\cite{PM14}, and in particular
for ensembles of theta-neurons~\cite{LBS13,SLB14,Lai14,Lai15,CB19,RM16}.    
The theta-neuron phase-model can be transformed
to a voltage-based description, the QIF model~\cite{EK86}.
Similarly, the macroscopic description for networks of
theta-neurons (in terms of the Kuramoto order parameter)
transforms into a more natural description
for ensembles of QIF neurons in terms
of two mean-field quantities of particular relevance in neuroscience:
the mean firing rate, and the mean membrane potential~\cite{MPR15}.

Such firing rate description for ensembles of QIF neurons
is remarkably simple and
amenable to mathematical analysis. This has motivated a
number of recent extensions
of the FRE for QIF neurons to a number of different setups
~\cite{RP16,DRM17,RP17,DEG17,SAM+18,ERA+17,Lai18,DVT18}.
In particular, considering the QIF model
in Sec.~\ref{sec:model}, the FRE consist of a system of 
two delay differential equations for the firing rate
$r$ and for the mean membrane potential
$$v= \int_{-\infty}^{\infty} d\eta \, g(\eta) \left[ \, \lim_{R\to\infty} \, \int_{-R}^{R} dV \, \rho(V|\eta,t) V\right] ,$$
which read~\cite{MPR15,PM16}
\begin{subequations}
\label{fre0}
\begin{eqnarray}
\tau \dot r &=& \frac{\Delta}{\pi\tau} + 2  r v,  \label{frer0}\\
\tau \dot v &=&   v^2 +\bar \eta -(\pi \tau  r)^2  +J \tau r_{D}. \label{frev0}
\end{eqnarray}
\end{subequations}
These FRE
describe the evolution of the population of infinitely many
spiking neurons in terms
of the firing rate $r$ and the mean-membrane potential $v$ of the population
of QIF neurons Eq.~\eqref{qif}.
Eqs.~\eqref{fre0} have 5 parameters, which can be
reduced to 3 by nondimensionalization.
In Ref.~\cite{PM16} the FRE Eqs.~\eqref{fre0} were analyzed under the
restriction $\bar \eta>0$, and they were rescaled
accordingly. Such rescaling allows to
systematically vary the time delay parameter $D$ (including the case 
$D=0$), and facilitates the
comparison with the classical and well-studied Kuramoto model with delay
~\cite{YS99,CKK+00,ES03,MPS06,LOA09}.

Alternatively, here we consider a new nondimensionalization
which allows us
to investigate the dynamics of the FRE Eqs.~\eqref{fre0} 
in the entire range of $\bar \eta$, so that the majority of the
neurons can be either self-oscillatory ($\bar \eta>0$) or quiescent/excitable
($\bar \eta<0$). Specifically, we rescale time and
$v$ by $D$ and $\tau$ as
\begin{equation} \label{rescal}
\tilde{t}=D^{-1} t \, , \quad \tilde{v}=D \tau^{-1} v, 
\end{equation}
so that the new, non-dimensional rate is $\tilde{r}=D r$.
Then the dynamics of the FRE can be completely explored, without
loss of generality, considering the rescaled parameters
$$
\tilde{J}=D\tau^{-1} J \, ,\quad \tilde{\bar\eta}=D^2 \tau^{-2} \bar{\eta} \, ,\quad \tilde{\Delta}=D^2 \tau^{-2} \Delta ,
$$
and setting $\tau=D=1$ in Eqs.~\eqref{fre0}. Specifically, we
investigate the nondimensional system of equations
\begin{subequations}
\label{fre1}
\begin{eqnarray}
\frac{d \tilde r}{d \tilde t} &=& \frac{\tilde \Delta}{\pi} + 2  \tilde r  \tilde  v,    \label{frer}  \\
\frac{d \tilde v}{d \tilde t} &=&    \tilde v^2 + \tilde{\bar \eta}
-(\pi   \tilde r)^2  + \tilde J   \tilde r_{D=1}.\label{frev}
\end{eqnarray}
\end{subequations}
To lighten the notation we drop the tildes hereafter (also in
the figure labels).

\section{Populations of Identical Neurons}

As we discussed previously,
the case of identical oscillatory neurons
has been investigated in~\cite{PM16} using 
a certain rescaling that required $\bar{\eta}>0$. 
Here we adopt the rescaling in Eq.~\eqref{rescal},
which allows us for an exhaustive investigation of the
dynamics of the system by systematically varying the parameter
$\bar \eta$.

Before starting the analysis, we emphasize that the Lorentzian ansatz
(or the equivalent OA ansatz)
is not strictly valid for identical oscillators.
In this case the system is partially integrable and its phase space is foliated by
a continuum of invariant manifolds, being the Lorentzian ansatz a particular one.
Actually, for the case of identical neurons ($\Delta=0$), the
correct approach is to resort to the so-called Watanabe-Strogatz theory
~\cite{WS94}, instead of the OA ansatz~\cite{PR11,Lai18}. Nevertheless,
from a physical perspective the OA/Lorentzian ansatz is very significant since
any small amount of noise and/or heterogeneity destroys the
degeneracy and, at least for the systems analyzed so far,
the density converges to a vicinity of the OA manifold~\cite{TGK+18}.

Hence, in the following we analyze the identical case
taking into account that its full significance holds once a small amount of
noise or heterogeneity is added to the system.
However, to avoid the inclusion of noise/heterogeneity
in the integration algorithm, we use initial conditions
in the Lorentzian manifold in all the numerical simulations
of ensembles of identical QIF neurons Eqs.~\eqref{qif}.

\subsection{Analytical results: The incoherent and the fully synchronized
states}

\subsubsection{The incoherent state}

\begin{figure}[t]
\centering
\includegraphics[width=90 mm]{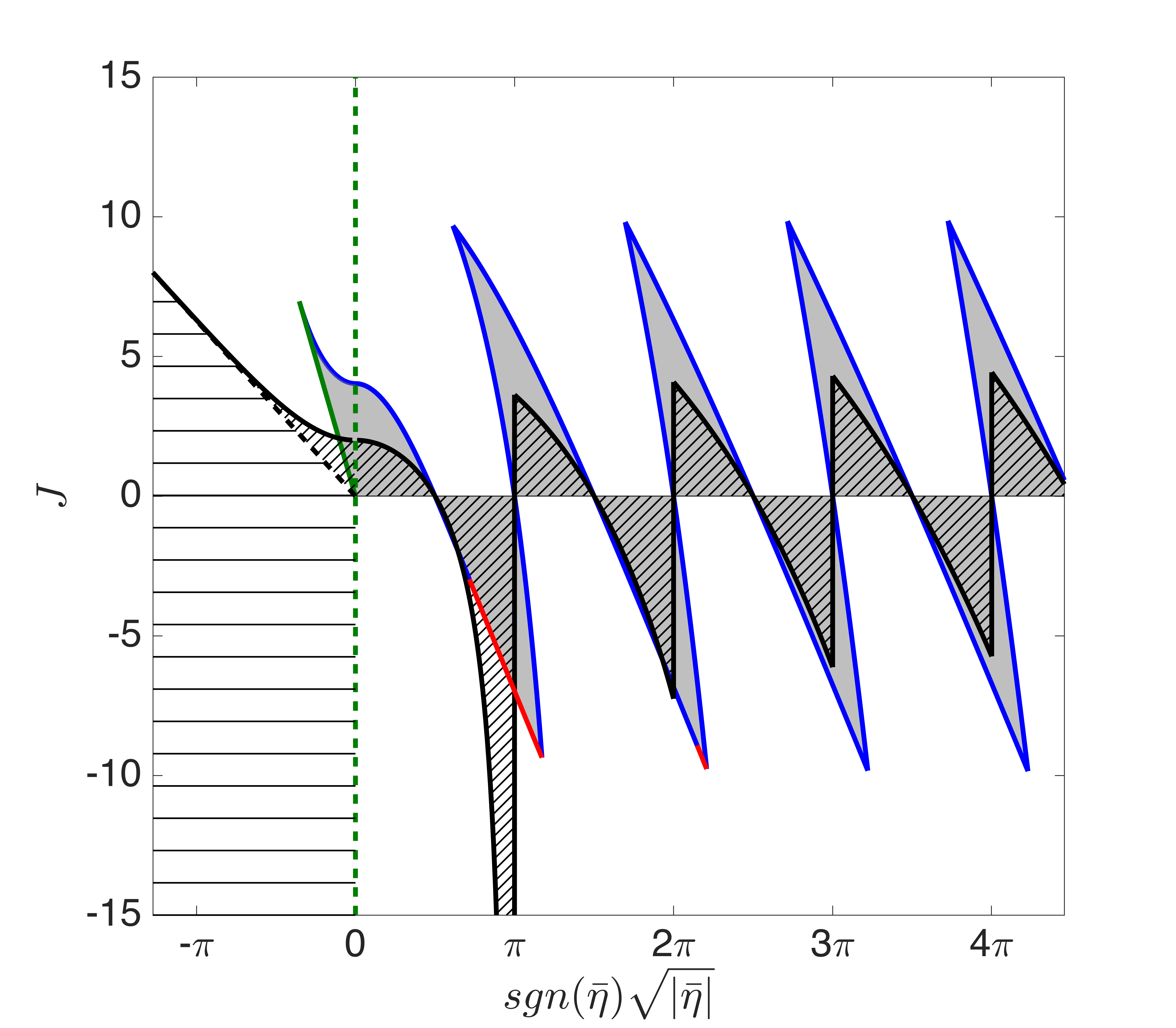}
\caption{Phase diagram for identical neurons, $\Delta=0$.
Shaded region: The asynchronous state
($\mathbf{a_{+}}$) is stable.
Slantwise hatched region: full synchrony is unstable.
Horizontally hatched region: The fully synchronized state does
not exist and the only attractor is the global rest state
$\mathbf{q}_-$.
The orbit of fully synchronized self-sustained
oscillations is created
at the dashed black line
(at $\bar{\eta}<0$), Eq.~\eqref{eq=synch_exist}.
Blue and red lines
are the loci of the sub- and super-critical Hopf-like
instabilities of incoherence Eqs.~\eqref{eq=Hopfs}.
Solid green line: saddle-node bifurcation. The vertical dashed green
line separates the oscillatory from the excitable regime of the QIF neuron.
}
\label{Fig1}
\end{figure}

Equation (\ref{fre1}) has at most four fixed points. In some parameter values
one of these points is located in the negative rate region ($r<0$),
and we refer to it as ``unphysical''.
Moreover, for $\Delta=0$,
the axis $r=0$ is invariant so that solutions initiated with $r(0)>0$
remain positive for all times.
The equilibria of Eqs.~\eqref{fre1} can be grouped into two sets of fixed points:
\begin{itemize}
\item The first pair of fixed points is located in the $(r,v)$ plane at
$$
\mathbf{a}_{\pm}=
\left(\dfrac{J\pm\sqrt{J^2+4\pi^2\bar{\eta}}}{2\pi^2},0\right) .
$$
For $J>0$, these fixed points are born
at a saddle-node bifurcation located at
\begin{equation}
J_\mathrm{sn}=2\pi\sqrt{-\bar\eta}.
\nonumber
\end{equation}
This line is partly depicted as a solid green straight line in the
phase diagram Fig.~\ref{Fig1}, and is located in the
region $\bar\eta<0$. Note that
the fixed point $\mathbf{a_{-}}$
becomes unphysical for $\bar\eta>0$, while
$\mathbf{a_{+}}$ exists for $J<0$ only if $\bar\eta>0$.
As shown below, the fixed point $\mathbf{a_{+}}$ is stable in a
wide range of parameter values. We will refer to $\mathbf{a_{+}}$
as the incoherent, or the asynchronous state.
For finite networks $\mathbf{a_{+}}$  becomes a so-called
splay state, with all neurons firing with the same ISI,
and one neuron firing every ISI$/N$ time units.

\item The second pair of fixed points,
$$
\mathbf{q}_{\pm}=\left(0,\pm\sqrt{-\bar{\eta}}\right),
$$
only exists for $\bar\eta<0$. They correspond to quiescent states,
and coincide with the fixed points of an
individual QIF neuron.
Hence, $\mathbf{q}_-$ (resp. $\mathbf{q}_+$) is trivially stable (unstable).
The bifurcation at $\bar\eta=0$ (green dashed line in
Fig.~\ref{Fig1})
is somewhat peculiar because it is not a simple saddle-node
bifurcation of $\mathbf{q}_{+}$ and
$\mathbf{q}_{-}$ as expected. For $J>0$, it involves the simultaneous
collision with $\mathbf{a}_{-}$, while for
$J<0$ it coincides with the appearance of $\mathbf{a}_{+}$ for $\bar\eta>0$.
\end{itemize}

Next we study the linear stability of the fixed points.
The incoherent state $\mathbf{a_{-}}$ is
always unstable, while the linear stability analysis of the high activity, asynchronous state $\mathbf{a_{+}}$ reveals interesting features.
Imposing the condition of marginal stability $\lambda=i\Omega$ in the
linearization of Eq.~\eqref{fre1},
we find a family of oscillatory instabilities at
\begin{equation}
J_{H}^{(n)}= \pi \left(\Omega_{n}^2-4\bar{\eta}\right)
\times
\begin{cases}
\left(6\Omega_{n}^2+12\bar{\eta}\right)^{-1/2}, ~\mbox{odd $n$}\\
\left(2\Omega_{n}^2-4\bar{\eta}\right)^{-1/2},  ~ \mbox{even $n$}  
\end{cases}
\label{eq=Hopfs}
\end{equation}
where $\Omega_{n}=n\pi$. We point out that these instabilities
(represented as blue and red lines in the phase diagram Fig.~\ref{Fig1})
are actually Hopf-like, rather than Hopf, because of two facts:
(i) The amplitude equations, computed in the Supplemental Material of \cite{PM16},
are degenerated.
(ii) In the supercritical case, we find that the emerging
limit cycle has a period $2\pi/\Omega_n$, which remains
constant as one moves away from threshold. This is apparently related to the reversible character of Eqs.~\eqref{fre1}
for $\tilde{\Delta}=0$ (note the invariance $t\to-t$, $v\to-v$) that,
as argued in \cite{PM16}, stabilizes symmetric orbits with
fixed periods when $D$ is nonzero.

 \setlength{\tabcolsep}{8pt}
\begin{table*} [t]
\centering 
\begin{tabular}{c  c c  c  c  c c } 
\cline{2-7}\cline{2-7}
 \hline\hline
 & &  \textbf{ASYNC} & \textbf{FULL SYNC} &  \textbf{QPS}  & \textbf{M-QPS} & \textbf{COLLECTIVE CHAOS} \\
\multirow{2}{*}{\textbf{IDENTICAL}} &  \textbf{Single neuron:} & Periodic & Periodic & 2F-Quasip. & 3F-Quasip. & Chaotic-like $(\lambda=0)$  \\  
& \textbf{Mean field:} & Constant& Periodic & Periodic & 2F-Quasip. & Chaotic \\[3pt] 
\hline
\rule{0pt}{10 pt}   
& &  \textbf{ASYNC}  &  \textbf{PS-I} &  \textbf{PS-II} &  \textbf{M-PS} &  \textbf{COLLECTIVE CHAOS} \\ 
\multirow{3}{*}{\textbf{HETEROGENEOUS}}
& \multirow{2}{*}{\textbf{Single neuron:}} &  \multirow{2}{*}{Periodic}   &  Periodic,&  Periodic, &  2F-Quasip.,&  \multirow{2}{*}{Chaotic-like $(\lambda <0)$}  \\
& & &2F-Quasip. &2F-Quasip. &3F-Quasip. & \\
& \textbf{Mean field:} &  Constant & Periodic & Periodic & 2F-Quasip. & Chaotic \\
\hline\hline 
\end{tabular}
\caption{Classification of the different dynamical states observed for
populations of both identical, and heterogeneous QIF neurons. 
The names of the states are the following: ASYNC: Asynchronous or incoherent state. FULL SYNC: fully synchronized state. QPS: Quasiperiodic partial synchronization. M
-QPS: modulated quasiperiodic partial synchronization.
PS-I and PS-II stands for type I and type II partially synchronous states.
M-PS: modulated partially synchronous state.
The prefix 2F- and 3F- indicate the number of frequencies of the corresponding
quasiperiodic dynamics.
For each state we specify the dynamics at the macroscopic level (mean field)
and at the microscopic level (single neuron). For the states of 
collective chaos, $\lambda$ is the Lyapunov 
exponent of a single neuron forced by the mean field. 
}
\label{table} 
\end{table*}

\subsubsection{The fully synchronized state}

Besides the stability boundary of the asynchronous state,
we can also analytically determine the boundaries of
full synchrony, $V_{j}=V(t), ~\forall j$.
The FRE Eq.~\eqref{fre1} are not suitable for this analysis,
since the fully synchronized
state corresponds to a degenerate infinite trajectory along the $v$-axis.
Full synchrony is hence investigated using the original
Eqs.~\eqref{qif}.

As shown in~\cite{PM16}, for oscillatory dynamics
($\bar\eta>0$) the stability region of full synchrony is  bounded
by the family of curves
\begin{equation}\label{eq=synch_identical_positive}
J_{s}^{(n')}=2\sqrt{\bar{\eta}}\cot \left({\dfrac{\sqrt{\bar{\eta}}}{n^{\prime}}}
\right) \quad \text{with} \quad n^{\prime}=1,3,..,
\end{equation}
and by the evenly spaced lines $\sqrt{\bar{\eta}}=m\pi$ with $m=1,2,3,...$.

On the other hand, in the case $\bar\eta<0$,
we emphasize that the term `full synchronization' cannot be strictly
used since the neurons are excitable and not self-sustained oscillators.
However, to simplify the notation,
in the following we refer to collective
oscillatory states with $\bar\eta<0$ as fully synchronized states.
Indeed, due to the presence of time delay, collective self-sustained
oscillations could be in principle maintained for strong enough
excitatory coupling.
To study the existence and stability of these states,
we rewrite the QIF model Eq.~\eqref{qif} as
\begin{equation}
\dot{V}_{j}=V_{j}^2-\vert\bar{\eta}\vert+Jr_{D}.
\label{qif_eta}
\end{equation}
Then, to investigate the existence of a fully synchronized state,
we can drop the index $j$ in Eq.~\eqref{qif_eta}. Note that,
in the absence of coupling,
Eq.~\eqref{qif_eta} has one stable ($s$) and one
unstable ($u$) fixed points
\begin{equation*}
V^*_u=-V^*_s=\sqrt{|\eta|}.
\end{equation*}
Between consecutive spikes, the evolution of the membrane potential of all
neurons is given by
\begin{equation}
\dot{V}=V^2-\vert\bar{\eta}\vert.
\label{eq=qif_ev}
\end{equation}
Considering that the neurons' membrane potentials reach
infinity at $t=0$,
we find that their membrane potentials at the time
immediately before receiving the spike, $t=D^-=1^-$, must satisfy the
following equation,
\begin{equation*}
\int_{-\infty}^{V\left(1^-\right)}\dfrac{dV}{V^2-\vert\bar{\eta}\vert}=1,
\end{equation*}
which gives
\begin{equation*}
V\left(1^-\right)\equiv V^{-}=-\sqrt{|\bar\eta|} \coth\sqrt{|\bar\eta|} .
\end{equation*}
A necessary condition for the existence of self-sustained collective oscillations
is that an excitatory spike causes a jump in $V$ beyond the unstable fixed point,
which enables the repetition of the cycle.
More precisely, immediately after receiving the first spike, $t=1^+$,
the membrane potential $V^+$ must satisfy $V^{+}>V^*_u$.
Then, for $\bar \eta<0$, we find that fully synchronized solutions
exist above the critical coupling
\begin{equation}
J_c=V^*_u-V^{-}=2 \sqrt{|\bar\eta|}\dfrac{e^{2\sqrt{|\bar\eta|}}}{e^{2\sqrt{|\bar\eta|}}-1}.
\label{eq=synch_exist}
\end{equation}
To analyze the stability of full synchrony, we study the evolution of
an infinitesimal perturbation $\delta V$ of a single neuron membrane potential
away from the cluster formed by the rest of the population.
The perturbed neuron and the cluster before the incoming
spike evolve according to the flow given by Eq.~\eqref{eq=qif_ev}.
The multiplier of the linearized flow ($\dot{\delta V}=2V \delta V$)
is antisymmetric causing convergence for negative $V$,
and divergence for positive $V$.
Hence, to have a stable fully synchronous solution, the
neurons need to spend more time in the convergent region of
the flow than in the divergent one.
This holds if the instantaneous jump of the membrane potential due to
the incoming spike is large enough. 
Then the critical coupling corresponds to $V^+=|V^-|$,
i.e.~$J_s=2|V^-|$, or
\begin{equation}
J_s=2 \sqrt{|\bar\eta|} \coth\sqrt{|\bar\eta|}
\label{eq=full_synch}.
\end{equation}
This function is precisely the boundary in
Eq.~(\ref{eq=synch_identical_positive}) with $n^{\prime}=1$,
which extends to the negative $\bar\eta$ region,
since $\cot(i x)= -i \coth(x)$.
Note also that $J_s$ approaches $J_c$ as $\bar\eta\to-\infty$.

\begin{figure*}
\centering
\includegraphics[width=170mm]{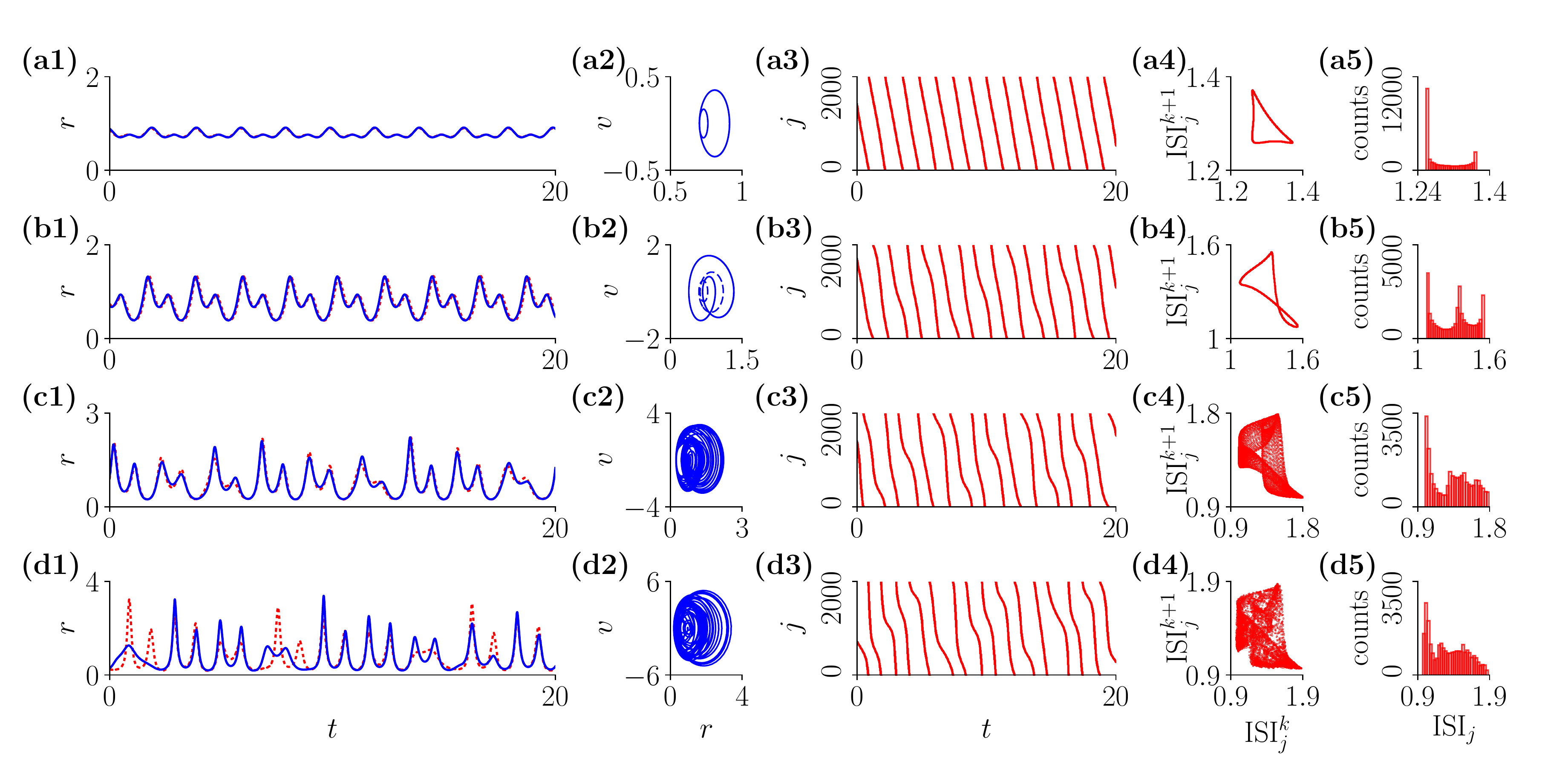}
\caption{Macroscopic (columns 1-2) and microscopic (columns 3-5)
dynamics of QPS (rows a,b), M-QPS (row c) and collective chaos (row d),
see Table~\ref{table}.
Column 1: time series of the mean firing rate. Blue lines
correspond to numerical simulations of the
FRE Eqs.~\eqref{fre1}, while red dotted lines are obtained computing the
mean firing rate of a population of $N=2000$ QIF neurons.
Column 2: $(r,v)$ phase portraits, numerically obtained using
Eqs.~\eqref{fre1}. In panel (b2) two coexisting
periodic attractors are shown: QPS-asym(I) (solid) and QPS-asym(II) (dashed)
---see also inset of Fig.~\ref{Fig3}. Panels (b1,b3-5) correspond to 
QPS-asym(I).
Columns (3-5) show the dynamics of a population of $N=2000$ QIF neurons.
Column 3: raster plots. Neurons are ordered according to their firing time at the beginning of the simulation 
(due to the first order nature of the QIF model, this ordering is preserved in time). Columns 4 and 5 show return ISI 
plots and ISI distributions of an arbitrary neuron $j$. 
The return plots of panels (a4,b4) are closed curves, indicating quasiperiodic microscopic dynamics in the QPS-sym and QPS-asym.
The corresponding ISI histograms (a5,b5) show two (QPS-sym) or three (QPS-asym) peaks. In the M-QPS, neurons are quasiperiodic with
three characteristic frequencies ---
the return plots of panel (c4) is a closed surface in 3D, and therefore its projection in 2D fills a defined region of the space. 
Parameters: (row a) $J=-9.2$, (row b) $J=-9.5$, (row c)
$J=-10.3$, (row d) $J=-10.6$. We use $\sqrt{\bar{\eta}}=3.6$ in all simulations.
}
\label{Fig2}
\end{figure*}

\subsection{Phase diagram}

The phase diagram shown in Fig.~\ref{Fig1}
summarizes our analytical results for populations of identical neurons. 
On the $y$ axis we represent the coupling strength $J$, which can be either excitatory or 
inhibitory. On the $x$ axis we represent a quantity that, if positive,
is proportional to the natural frequency of the neurons, see Eq.~\eqref{isi}.
Regions with qualitatively different dynamics are highlighted with different
combinations of colors and patterns.
In the gray shaded regions, the asynchronous state $\mathbf{a_{+}}$ is stable, while
slantwise hatching indicates instability of the fully synchronized state.
On the other hand, in the horizontally-hatched area,
the global quiescent state $\mathbf{q_{-}}$ is the only attractor of the system.
In the unhatched white region, full synchrony is a stable attractor
(and typically the only one--see below),
but several of such states may coexist in certain regions
for $\bar{\eta}>0$.

More specifically, in the excitable region ($\bar{\eta}<0$) of the diagram
the global quiescent state $\mathbf{q_{-}}$ is always stable.
In addition,
the stability region of the asynchronous state $\mathbf{a_{+}}$ (grey shading)
is bounded by the saddle-node bifurcation $J_{\text{sn}}$ (green line),
and the Hopf-like bifurcation line $J_H^{(1)}$, Eq.~\eqref{eq=Hopfs}
(blue line). The two lines meet at a Zero-Hopf codimension-two point.
In the unhatched grey region $\mathbf{a_{+}}$
coexists not only with $\mathbf{q_{-}}$,
but also with the fully synchronized state. This oscillatory
state becomes stable at the solid black line Eq.~\eqref{eq=full_synch}.

On the other hand, the positive $\bar{\eta}$ region of the
diagram is characterized by a sequence
of subcritical (blue lines) and supercritical (red lines) Hopf-like bifurcations,
defined by Eq.~\eqref{eq=Hopfs}, that switch the stability of the incoherent state
$\mathbf{a_{+}}$. Remarkably, in this region (where neurons are 
self-sustained oscillators), the phase diagram 
bears strong resemblance with that of the
Kuramoto model with time delays~\cite{YS99,CKK+00,ES03,MPS06,LOA09}.
The two systems display tent-shaped
regions with an even spacing given by the equality between the delay ($D=1$)
and the intrinsic ISI Eq.~\eqref{isi}, as well as
bistability regions between full sync and incoherence
(unhatched gray regions).
However, while in the Kuramoto model the Hopf
bifurcations are always subcritical, here we find supercritical
Hopf bifurcations for some $\bar \eta$ values in the inhibitory ($J<0$)
part of the diagram. Near the supercritical Hopf bifurcations, in the unshaded hatched region, both the incoherent and
the fully synchronous states are unstable, and partial
synchrony (QPS, M-QPS, and collective chaos) is found.
In the next section we classify these states in terms of their 
macroscopic and microscopic dynamics, and investigate their bifurcations.

Finally, we discuss an interesting feature of the phase diagram
in Fig.~\ref{Fig1} ---see also the phase diagram in \cite{PM16}.
Note that, at variance with the vertically oriented, tent-shaped 
regions of the Kuramoto model~\cite{YS99,CKK+00,ES03,MPS06,LOA09},
here the regions of stability are tilted.
This discrepancy between populations of QIF neurons and the Kuramoto model
can be understood as follows: in the QIF model the neurons always
advance their phase in response to excitatory inputs, and always delay their phase
in response to inhibitory inputs ---i.e. they have a so-called Type 1 phase
resetting curve.
This produces the progressive `advancement' of the boundaries in the
excitatory part of the phase diagram
as the strength of the excitatory coupling $J$ is increased
---given that neurons increase their
firing frequency and thus their effective value of $\bar \eta$.
Similarly, in the inhibitory region,
the neurons slow down their firing frequency in response to inhibitory inputs,
and this progressively `delays' the boundaries for $J<0$.
In contrast, in the classical Kuramoto model, the terms producing
advances and delays in response to excitation and inhibition are
not included~\cite{MP18}, and hence the boundaries are not tilted.

\subsection{Numerical analysis of partially synchronous states}

\begin{figure}
\centering
\includegraphics[width=75 mm]{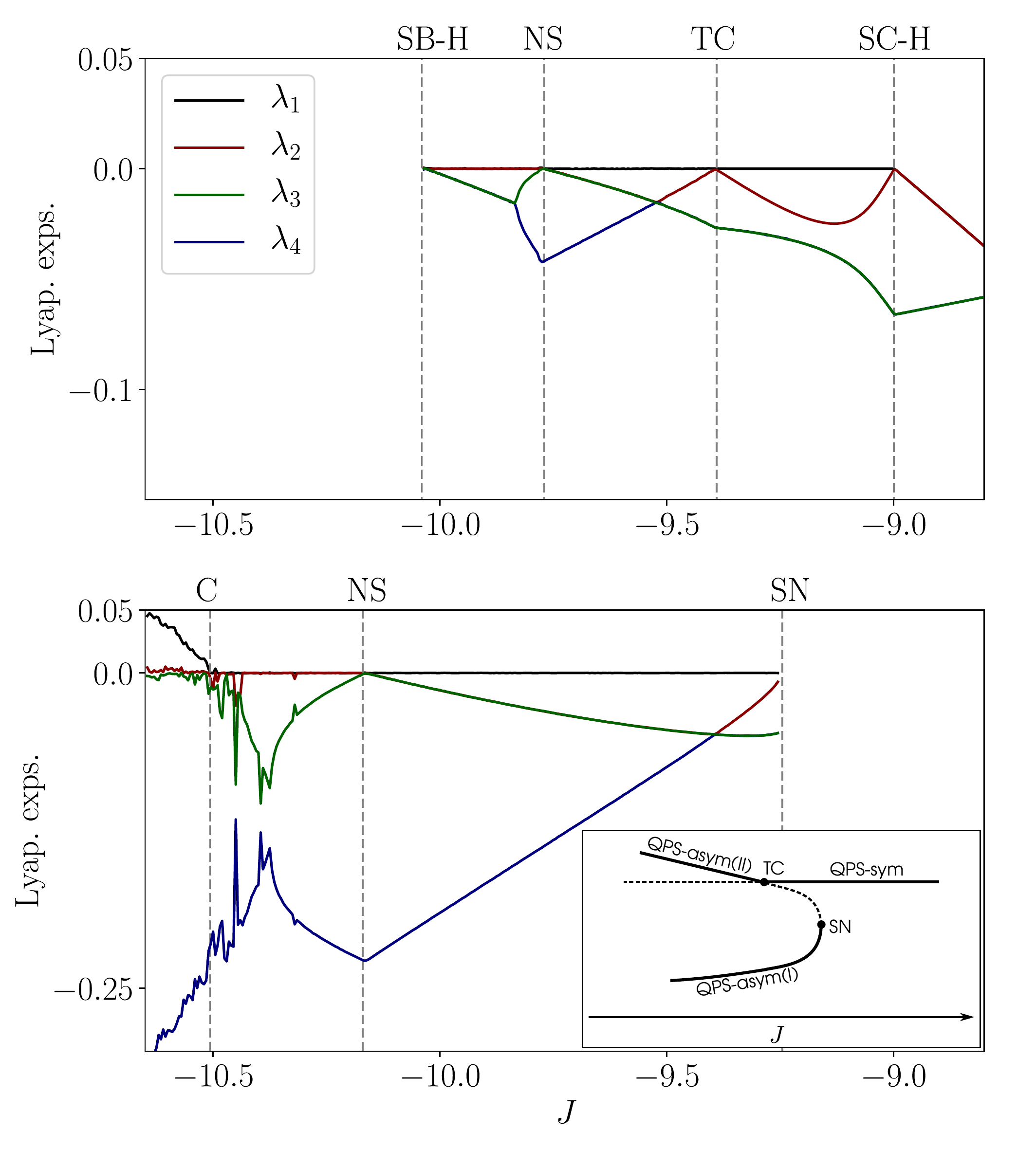}
\caption{Four largest Lyapunov exponents for two alternative bifurcation sequences in a range of negative $J$ values and
fixed $\sqrt{\bar{\eta}}=3.6$. For each solution, the continuation 
was carried out either increasing or decreasing parameter $J$ adiabatically.
In the top panel the vertical dashed lines indicate, 
from right to left: a supercritical Hopf bifurcation (SC-H),
a transcritical bifurcation (TC), a Neimark-Sacker bifurcation (NS), and a subcritical Hopf bifurcation (SB-H). In the bottom panel the vertical dashed lines indicate,
from right to left: a saddle-node bifurcation (SN), a Neimark-Sacker bifurcation (NS), and the onset of chaos (C).
The inset shows a sketch of the bifurcation diagram connecting the two bifurcation sequences.}
\label{Fig3}
\end{figure}

Next we perform a numerical exploration of the partially synchronized states
arising both in the white slantwise-hatched region of Fig.~\ref{Fig1}, as well 
as in some neighboring regions. 
In Table \ref{table} these partially synchronized states are classified 
according to their dynamics, both for identical and for
heterogeneous (in Sec.~V) populations of QIF neurons.
The macroscopic dynamics of the states is investigated performing 
numerical simulations of the FRE Eq.~\eqref{fre1}, and illustrated 
in the columns 1 and 2 of Fig.~\ref{Fig2}. To investigate the single neuron
dynamics associated to the macroscopic states we also 
performed numerical simulations of the original system of QIF neurons 
Eqs.~\eqref{qif}, and depicted the raster plots (column 3), and the 
ISI return maps (column 4) and histograms (column 5). Finally, in column 1, 
we also show the time series of the population-mean firing rate of the 
network simulations (dashed red lines), which show a perfect agreement with the
time series of the FRE (blue lines) ---except in panel 
(d1), where the collective dynamics is chaotic.

Note that stable partially synchronized states are not only found in the 
slantwise-hatched region of Fig.~\ref{Fig1}, but also in a 
neighborhood of this region with  $\sqrt{\bar \eta}>\pi$. 
This is because the region where 
the Hopf-like bifurcation $J_{H}^{(1)}$ is supercritical 
(around the red line at $\sqrt{\bar \eta} \approx \pi$ in Fig.~\ref{Fig1}) 
extends 
to $\sqrt{\bar \eta}>\pi$, and hence one expects 
a low-amplitude periodic solution
bifurcating from incoherence, $\mathbf{a_{+}}$, coexisting with a 
fully in-phase synchronized state.
In Figs.~\ref{Fig2}(a1) and ~\ref{Fig2}(a2) 
we respectively show the time series and 
the phase portraits corresponding to 
the numerical integration of Eqs.~\eqref{fre1} for $\sqrt{\bar \eta}=3.6$. 
These simulations  
confirm the presence of a small amplitude symmetric limit cycle, 
which grows in size as parameters are moved away from the instability.

As analyzed in \cite{PM16}, in Fig.~\ref{Fig2}(a1) the oscillation period
of the mean field is exactly
$T=2$ (or, in dimensional form, $T=2D$).
The periodic dynamics of the global quantities
leads to quasiperiodic dynamics of the individual neurons,
i.e.~Quasiperiodic partial synchrony (QPS). 
This may be appreciated plotting the ISIs of a single
neuron versus their consecutive ISIs. The resulting return plot, shown in 
Fig.~\ref{Fig2}(a4), forms a closed curve indicating quasiperiodic dynamics.
Interestingly, the ISIs of the neurons are always shorter than the
period of the firing rate oscillations, 
as shown by the ISI histogram in Fig.~\ref{Fig2}(a5).
The bimodal structure of the distribution is related to
double-loop shape of the macroscopic periodic attractor.

The limit cycle that emerges via the Hopf-like instability
displays a robust $v\to-v$ symmetry that only breaks down
after another bifurcation. In \cite{PM16}, for $\sqrt{\bar \eta}=3$, 
it was shown that symmetry
broke down after a period-doubling bifurcation. Here,
taking a slightly larger value of $\sqrt{\bar\eta}$ and    
increasing inhibition, we observe an imperfect symmetry breaking 
transition, with two coexisting attractors, 
see Fig.~\ref{Fig2}(b1,b2) and Fig.~\ref{Fig3}. 
These asymmetric periodic orbits ---which we call QPS-asym(I) and QPS-asym(II)---
are not related by symmetry.
In fact, each attractor is born via a different bifurcation, 
see details below.
In these asymmetric states the period differs from $2D$, but still neurons
are quasiperiodic, see Fig.~\ref{Fig2}(b4,b5).

Increasing inhibition further, the macroscopic
dynamics becomes more irregular, with no evident periodicity, see Fig.~\ref{Fig2}(c1,c2). Below, we show the analysis of the Lyapunov exponents indicating quasiperiodic mean field dynamics with two incommensurable frequencies.
As a consequence of this quasiperiodic forcing, the neurons exhibit
three-frequency quasiperiodic motion, see Fig.~\ref{Fig2}(c4).
We call this new state modulated QPS, or simply M-QPS, due to the
additional modulating frequency.
To the best of our knowledge this state has been only 
reported 
in a very different setup 
~\cite{NK95,clusella18}.
Lowering $J$ further, the M-QPS eventually turns into a chaotic state, see Fig.~\ref{Fig2}(d1,d2).

\begin{figure}
\centering
\includegraphics[width=75 mm]{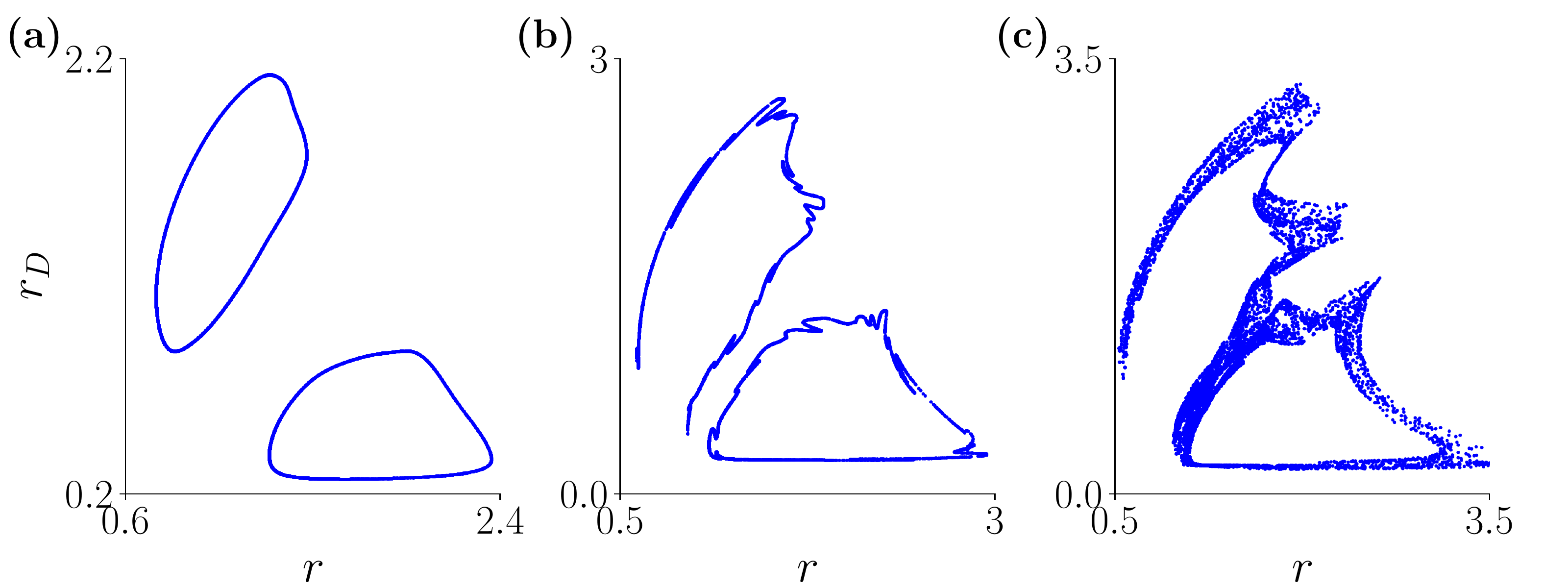}
\caption{Poincar\'e sections of the FRE \eqref{fre1} for
$\sqrt{\bar\eta}=3.6$, and for three different values of
the inhibitory coupling strength:
(a) $J=-10.3$; (b) $J=-10.48$; (c) $J=-10.6$.
The Poincar\'e surface is $v=0, ~\dot{v}<0$.}
\label{Fig4}
\end{figure}

To determine the bifurcations linking different partially synchronous
states (QPS, M-QPS, or collective chaos),
we computed the four largest Lyapunov exponents of the FRE
on the line along the $J$ direction with $\bar{\eta}$ value of Fig.~\ref{Fig2}.
Employing the usual method \cite{farmer82},
parameter $J$ was quasi-adiabatically decreased and increased, to detect eventual bistabilities.
Two parallel sequences of bifurcations were eventually detected,
as shown in top and bottom panels of Fig.~\ref{Fig3}.
In the top panel, moving leftwards, the fixed point attractor ($\mathbf{a}_+$),
first undergoes a supercritical Hopf-like bifurcation, after which the
stable attractor of the system is a symmetric QPS attractor.
The symmetry breaking takes place at a transcritical bifurcation (TC), after which the
limit cycle is asymmetric (QPS-asym(II)). At a lower $J$ value, the asymmetric periodic orbit
undergoes Neimark-Sacker bifurcation giving rise to M-QPS
---given that we find two vanishing Lyapunov exponents.
Further decreasing inhibition, the M-QPS disappears in
a subcritical Hopf bifurcation (SB-H).

\begin{figure}
\centering
\includegraphics[width=85 mm]{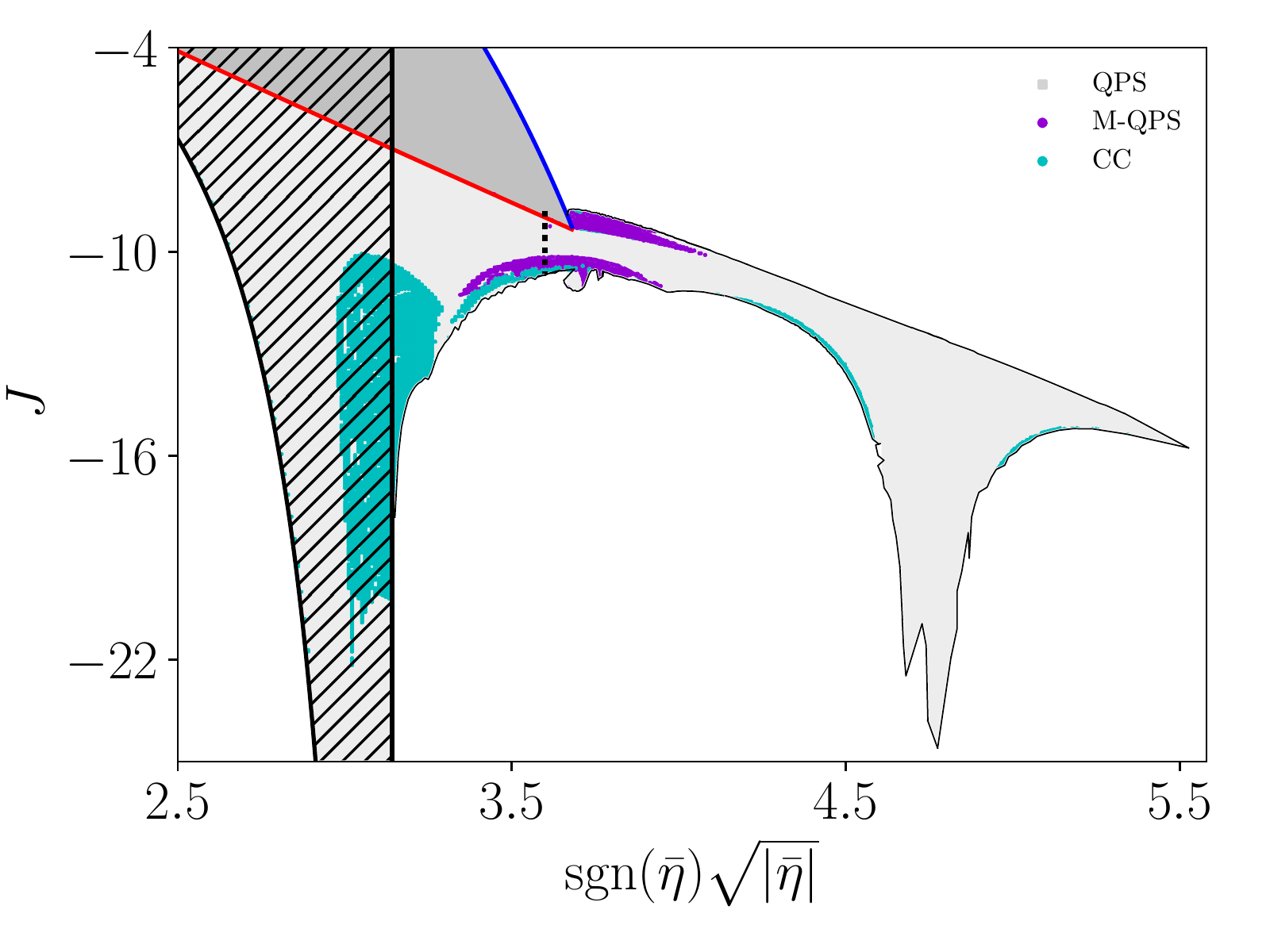}
\caption{Numerical exploration of the partially
synchronized states (QPS,M-QPS, collective chaos) near the
supercritical Hopf bifurcation in phase diagram Fig. \ref{Fig1}.
In the light gray region the largest Lyapunov exponent is zero, and
QPS is stable. The purple dots correspond to two vanishing Lyapunov exponents,
indicating quasiperiodic dynamics.
In the cyan region the dynamics is chaotic. The vertical dashed black line
at $\sqrt{\bar\eta}= 3.6$ corresponds to the range of parameters
explored in Fig.~\ref{Fig3}.}
\label{Fig5}
\end{figure}

In the other sequence of bifurcations ---bottom panel of Fig.~\ref{Fig3}--- 
another asymmetric orbit (QPS-asym(I))
is born at a saddle-node (SN) bifurcation.
As QPS-asym(II), it also undergoes a Neimark-Sacker bifurcation as $J$ is decreased giving rise to M-QPS.
In Figs.~\ref{Fig2} and \ref{Fig4} 
we show the M-QPS state corresponding to this particular sequence of bifurcations. 
However, note that M-QPS states resulting from either route 
in Fig.~\ref{Fig3}
have the same dynamical features (two vanishing largest Lyapunov exponents
and three-frequency microscopic motion).
Lockings occur at certain windows of $J$, where the second largest Lyapunov exponent is not zero.
To further prove the macroscopic quasiperiodic nature of the M-QPS, we also show
Poincar\'e sections for three different values of $J$ in Fig.~\ref{Fig4}. As $J$ is lowered
the torus corrugates as typically observed in the transition
to chaos via fractalization of the torus \cite{curryorke}, see Fig.~\ref{Fig4}(b).
The torus breaks down around $J=-10.5$, and 
the attractor turns fractal. Notably, the chaotic attractor
achieves rapidly an information dimension larger than three
according to the Kaplan-Yorke formula \cite{kapyor} since
$\lambda_1>\vert \lambda_3\vert$, see bottom panel of Fig.~\ref{Fig3};
in contrast with the dimension slightly above two found in \cite{PM16} for the chaotic attractor
born from the period doubling cascade. 
It is important to stress that, in spite of the
positive Lyapunov exponent (of the collective dynamics), 
the microscopic dynamics
remains nonchaotic, because the individual oscillators have only one degree of freedom.
In fact the structure of the model imposes the neurons to fire sequentially, see Fig.~\ref{Fig2}(d3).
Finally, the inset in Fig.~\ref{Fig3} is our conjecture of how the two bifurcation
sequences in the main panels are connected: the unstable branch the SN bifurcation collides with the
symmetric QPS state at the TC bifurcation.

In the preceding figures we have shown the transitions along a specific $\bar\eta$ value. Seeking a more global picture
we decided to sweep parameters $J$ and $\bar\eta$ monitoring
the largest Lyapunov exponents.
This permits to identify the attractor types efficiently. Figure~\ref{Fig5}
shows the region spanned by partially synchronized dynamics
~\footnote{Actually, we have not explored the region
close to the supercritical bifurcation just above $\sqrt{\bar\eta}=2\pi$.}.
The light gray and purple regions indicate
QPS and ~M-QPS states, respectively, while cyan dots correspond to chaotic dynamics.
It surprised us the extension of the parameter region where QPS coexists with full synchrony (light shaded unhatched area).
There is a ``tongue'' extending to very negative $J$ values around
$\sqrt{\bar\eta}=4.7$ that looks like an ``echo'' at $3\pi/2=4.712\ldots$
of the infinite tongue just below $\sqrt{\bar\eta}=\pi$. We have not an intuitive explanation for this.
Quasiperiodic dynamics (M-QPS) is found always not far from the degenerate point were the instability boundaries
for $n=1$ and $n=2$, see Eq.~\eqref{eq=Hopfs}, meet. This is probably not casual (further analysis
is nonetheless beyond our scope\footnote{The degenerate
point is a codimension-three point because the instability for $n=1$ is degenerate exactly at that point (see the Supplemental Material of \cite{PM16}).}). 
 Regarding the chaotic state, it shows up in two distinct regions: the leftmost one 
is related to the period-doubling scenario observed in \cite{PM16}, while the rightmost one is correspond to the
quasiperiodic route uncovered here.

\begin{figure}
\centering
\includegraphics[width=75mm]{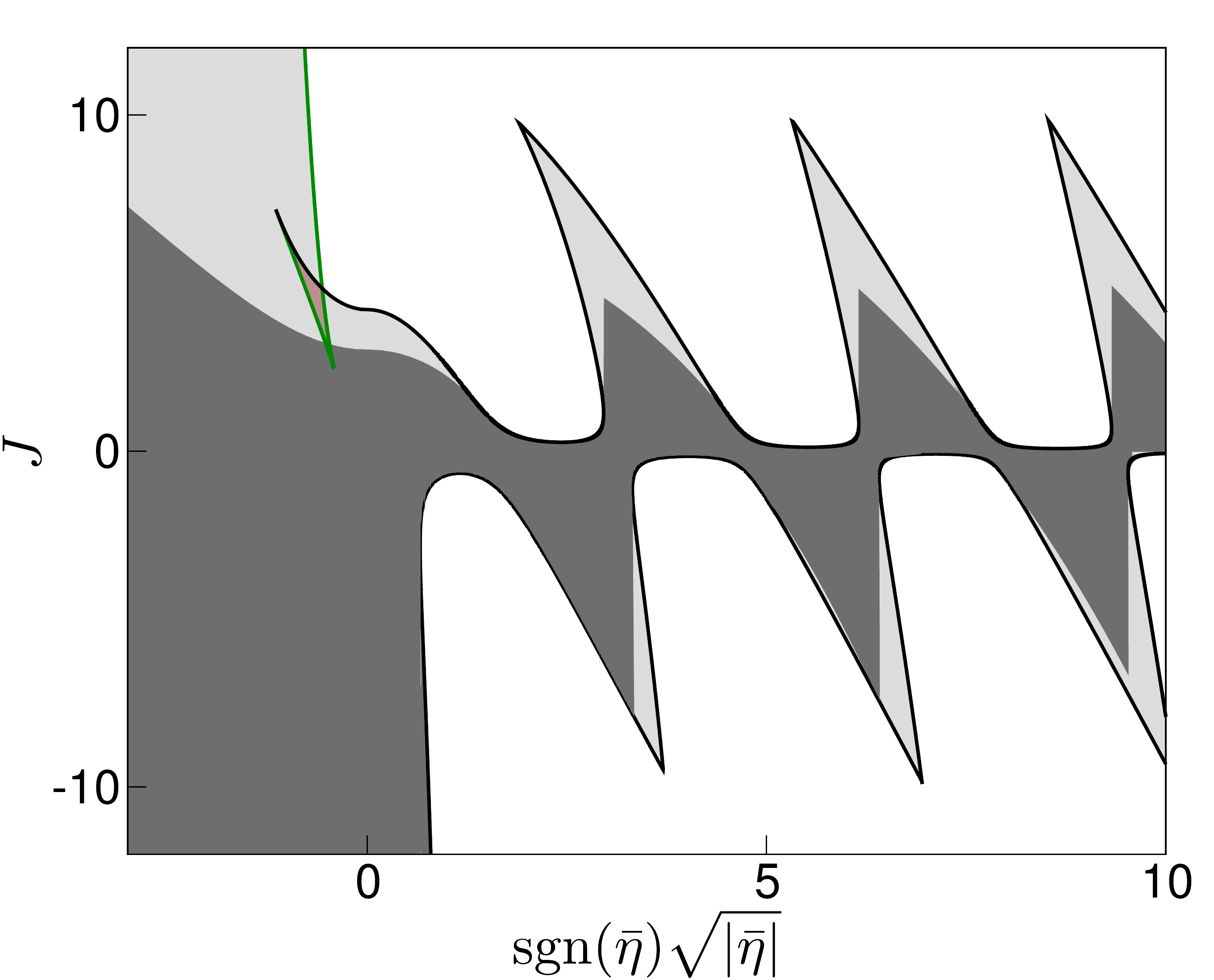} \hfill
\caption{
Phase diagram for populations of heterogeneous neurons, $\Delta=0.1$.
Dark shaded region: Incoherence (fixed point) is the only stable state.
Light shaded region: Incoherence (fixed point) coexist with a partially synchronous state (limit cycle). 
Brown region: Two forms of asynchrony (high and a low activity fixed points) coexist with a partially synchronous state.
Green lines are saddle-node bifurcations, and
black lines correspond to Hopf boundaries. Note that here, in contrast with Fig.~\ref{Fig1}, the Hopf boundaries are not represented in Blue/Red (we do not explicitly specify whether these boundaries are subcritical or supercritical).
The boundary between light and dark shaded regions was obtained numerically.}
\label{Fig6}
\end{figure}

\section{Populations of Heterogeneous neurons}

\begin{figure*}
\centering
\includegraphics[width=160 mm]{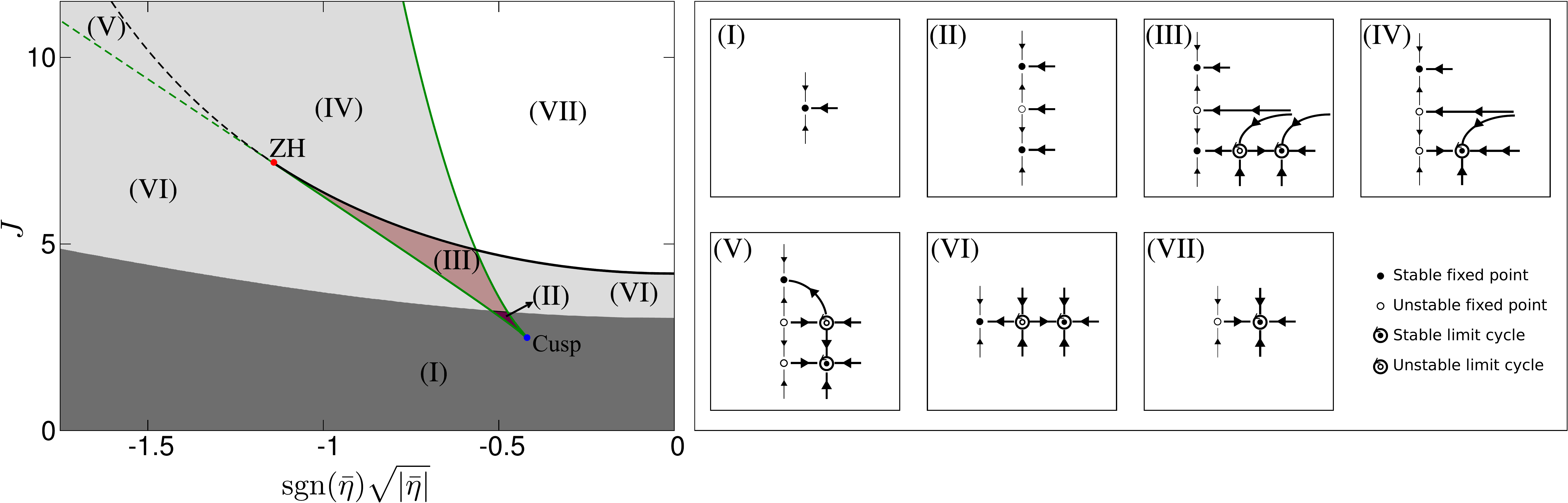} \hfill
\caption{Enlarged view of the region of
multistability located at $\bar{\eta}<0$ in Fig.~\ref{Fig6}.
Black line: Hopf bifurcation (subcritical). Green lines:
saddle-node bifurcations. In the dark shaded region, only the quiescent, low
activity state is stable. In the light shaded region, incoherence coexists with
a collective oscillatory state ---self-sustained due to recurrent excitation.
In the brown region the low activity fixed point coexists with
a high activity fixed point and with the oscillatory state.
In the small dark purple region only the two high and low
activity fixed points are attracting.
Right panels: Sketches of the Poincar\'e section
in different regions (assuming that
it coincides with the one-dimensional manifold that
connects different fixed points).
The thick lines indicate two-dimensional manifolds,
and periodic orbits are indicated by a point surrounded by  a small circle.}
\label{Fig7}
\end{figure*}

In this section we consider that the neurons in the network are non-identical,
and investigate how this alters the phase diagram in Fig.~\ref{Fig1},
and the partially synchronous states depicted in Fig.~\ref{Fig2}.
Hence, in the following we assume that the half-width $\Delta$ of the
Lorentzian distribution Eq.~\eqref{eq=lorentzian} is not zero.
Under the presence of Lorentzian heterogeneity fully and 
partially synchronous states discussed previously are unattainable.
In the following the generic term
`partial synchronization' refers to any state of the network
which is not an incoherent state.

States reminiscent of QPS and collective chaos persist for
finite values of $\Delta$, with individual neurons displaying different motions depending on their native $T_j$ values.
We denote these states as partial synchronization-I (PS-I)
and PS-II for the states reminiscent of full synchrony and QPS, respectively.
In PS-I most neurons are 1:1 entrained to the global frequency,
and the remaining neurons are either entrained with a different ratio
or display quasiperiodic dynamics. In the case of PS-II only a
minority of the neurons entrains 1:1 with the
macroscopic oscillation.
We use the distinction between PS-I and PS-II
for convenience, but we emphasize that there is not
a clearcut distinction between both states and one can transit
from one to the other continuously.
As for the other states, the asynchronous state continues
to exist after introducing the heterogeneity, although
not in the form of a splay state.
Finally, M-QPS is replaced by a modulated PS states,
or M-PS, while collective chaos continues to exist, see Table I.

Next we analyze how the stability regions of incoherence, which can still be 
analytically computed from the FRE Eqs.~\eqref{fre1}, change due to the presence
of heterogeneity. Unfortunately, in the heterogeneous case, a stability 
analysis similar to that of Sec. IVA for the case of synchronous states
is not possible. 
Later in this section we examine how the partially synchronized states found in the region
$\bar \eta>0$ for identical inhibitory neurons are altered by quenched disorder.

\subsection{Stability boundaries of incoherence and phase diagram for $\Delta=0.1$}

It is important to note that the presence of
heterogeneity removes all degeneracies of the FRE
Eqs.~\eqref{fre1}. The fixed points can be still obtained
in parametric form, as well as the boundaries corresponding to
saddle-node bifurcations of the asynchronous/incoherent states,
[green lines in Fig.~\ref{Fig6}].
However, these expressions are lengthy
and here we omit them for the sake of clarity, see~\cite{MPR15}.
Linearizing and imposing
the condition for marginal stability, also the loci of the Hopf
bifurcations can be
obtained in parametric form [black lines in Fig.~\ref{Fig6}].
We finally used numerical simulations of
Eqs.~\eqref{fre1} to detect the regions where partially synchronous
states become unstable, or cease to exist
[dark gray region in Fig.~\ref{Fig6}].

The phase diagram in Figure~\ref{Fig6} summarizes these results for $\Delta=0.1$,
and displays the regions where distinct dynamics occur ---compare with
the phase diagram Fig.~\ref{Fig1}. As expected, close to the $J=0$ axis
incoherence is the only attractor of the system (dark shaded).
Like in the case of identical neurons,
bistability regions between incoherence and another state(s) exist (light shaded).
Interestingly, for inhibitory coupling,
the Hopf bifurcations of the asynchronous state
largely overlap with the numerical boundaries of `pure' incoherence (dark
shading). This indicates that, for inhibitory networks,
the intervals where the Hopf bifurcations
are supercritical are dramatically enlarged as heterogeneity is increased.

\begin{figure*}[t]
\centering
\includegraphics[width=170 mm]{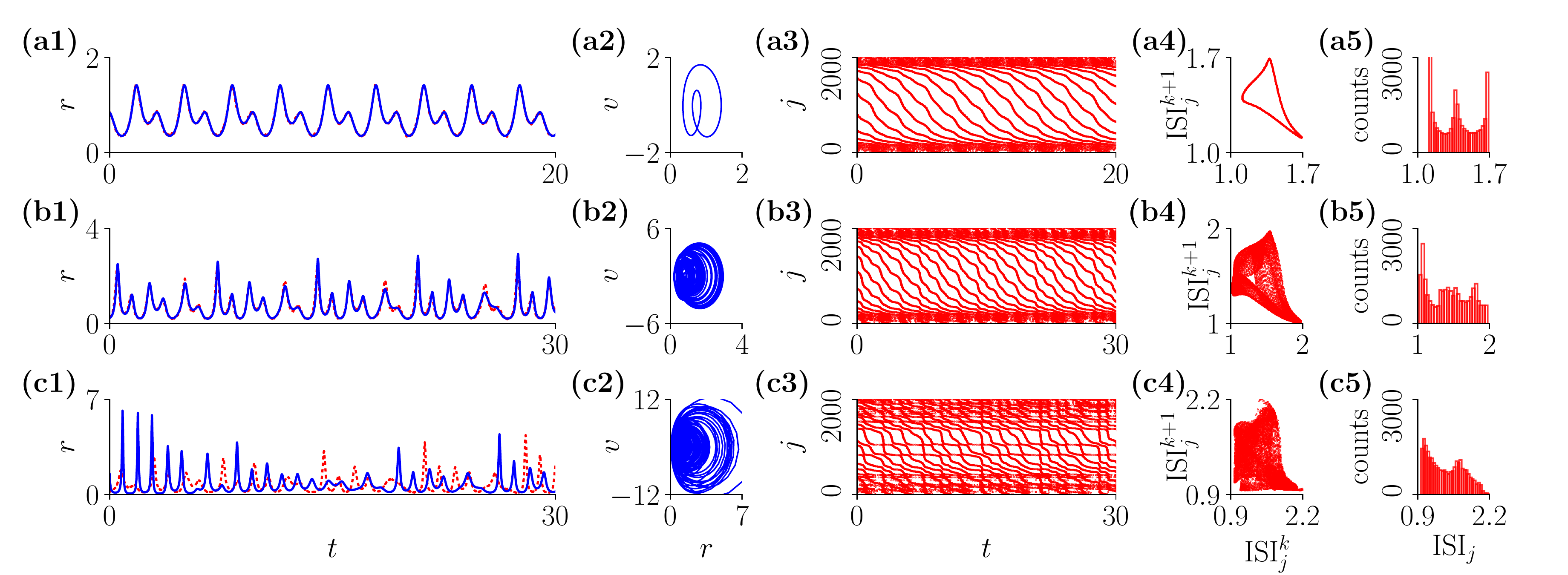}
\caption{Macroscopic (columns 1-2) and microscopic (columns 3-5)
dynamics of (row a) PS-II states, (row b) M-PS states, and
(row c) collective chaos for heterogeneous neurons,
---see Fig.~\ref{Fig2} and Table \ref{table}.
Column 1: Time series of the firing rate for the FRE Eqs.~\eqref{fre1} (blue) and for a population of $N=2000$ QIF neurons Eqs~.\eqref{qif} (red dotted).
Column 2 shows the corresponding attractors, obtained using the FRE.
In rows (a) and (b), the dynamics is periodic
but, in contrast with the identical case, here
the limit cycle is asymmetric due to the presence heterogeneity.
Column 3 shows the raster plots corresponding to numerical simulations
of a population of $N=2000$ QIF neurons Eqs.~\eqref{qif}, and columns 4 and
5 show the corresponding return plots and ISI histograms, respectively.
In the raster plots, each neuron index $j$ corresponds to a specific $\eta_j$ value (see text).
For the computation of
return plots and ISI histograms we used neuron $j=500$.
In panels (a4) and (b4) one can see that the neuron behaves
quasiperiodically, with
two and three incommensurable frequencies, respectively.
Note also the three peaks in panel (a5) due to the asymmetry
of the limit cycle.
In all panels we use $\Delta = 0.1$, $\sqrt{\bar{\eta}}=3.5$,
and (row a) $J=-9.60$; (row b) $J=-10.70$; (row c) $J=-11.30$.
}
\label{Fig8}
\end{figure*}

\subsubsection{Phase diagram in the region $\bar\eta<0$}

Figure~\ref{Fig7} displays an enlarged view of the phase diagram Fig.~\ref{Fig6},
around the brown region located at $\bar\eta<0$. The scenario of bifurcations
is quite intricate in this region, and here we describe it in detail. 
The brown shaded region is interesting since a high-rate
and a low-rate fixed points ---reminiscent of the fixed points
$\mathbf{a_{+}}$ and $\mathbf{q_{-}}$---
coexist with a periodic orbit. In Fig.~\ref{Fig7}
we have included two dashed lines corresponding to bifurcations involving
saddles and/or repellors to fully clarify the transitions between different
stable states. We also highlight two codimension-two points:
the cusp point where the two saddle-node
bifurcations meet, and the zero-Hopf (ZH) point ---associated to a zero and
a pair of pure imaginary eigenvalues. The different shadings in the figure
indicate regions with qualitatively different attractors:
in the dark region (I) only one fixed point is stable. In
the small dark purple region (II) this fixed point coexists
with another stable fixed point.
In the light shaded areas (IV,V,VI) a stable fixed point
coexists with a stable limit cycle. This limit cycle is the only
stable attractor in the white region (VII). Finally,
in the brown region (III), there are three coexisting
stable attractors: two fixed points, and a limit cycle.

The transitions between any two regions in the diagram can be
understood considering a three-dimensional space.
In the right panels of Fig.~\ref{Fig7}
we present sketches of the  phase portraits of the different stability regions,
by means of Poincar\'e sections. Thick lines represent two-dimensional manifolds.
Comparing the phase diagram in Fig.~\ref{Fig7} with the results
previously obtained for instantaneous interactions~\cite{MPR15},
we see that the delay promotes the appearance of a
 Hopf bifurcation of the asynchronous state.
Note that the scenario shown in Fig.~\ref{Fig7} resembles that of a
population of heterogeneous QIF neurons
with fast synaptic kinetics~\cite{RP16}, but here
we find a codimension-two ZH point, instead of a double-zero eigenvalue point.

\subsection{Numerical analysis of partially synchronized
states in the presence of heterogeneity}

Here we explore numerically how the presence of heterogeneity transforms the
partially synchronized states described in Sec.~IV. 
In order to circumvent sample-to-sample fluctuations,
$\eta_j$ values are selected deterministically from the Lorentzian distribution 
setting $\eta_j=\bar{\eta}+\Delta \tan\left[\pi(2j-N-1)/(2N+2)\right]$, where $j=1,2,\ldots,N$.
States reminiscent of previous 
partially synchronous states
persist for $\Delta \neq 0$; in columns (1,2) of Fig.~\ref{Fig8} we show the macroscopic 
time series of PS-II, M-PS and collective chaos, 
where blue lines represent
numerical integration of the FRE \eqref{fre1} and red lines simulation of a
population of QIF neurons. All the three states are clearly reminiscent of the
QPS, M-QPS and collective chaos states for identical neurons. 
In the columns (3-5) of Fig.~\ref{Fig8} we also show the raster plots of the spiking activity
of the population of QIF neurons together with the return plots and ISI
histograms of a single neuron of the population. 
Due to the presence of heterogeneity, in the PS-II state neurons can be either
periodic or two-frequency quasiperiodic, while in the M-PS they can be two- or
three-frequency quasiperiodic, see Table~\ref{table}. 
The illustrative neuron chosen to plot the
return maps and ISI histograms of Fig.~\ref{Fig8} are, respectively,
two-frequency and three-frequency quasiperiodic for panels (a4,a5) and (b4,b5).
Note how, as in the QPS-asym in Fig.~\ref{Fig2}(f), the histogram of ISIs for a quasiperiodic neuron in the
PS-II state has three peaks, due to the asymmetric shape of the limit cycle. 

\begin{figure}
\centering
\includegraphics[width=85mm]{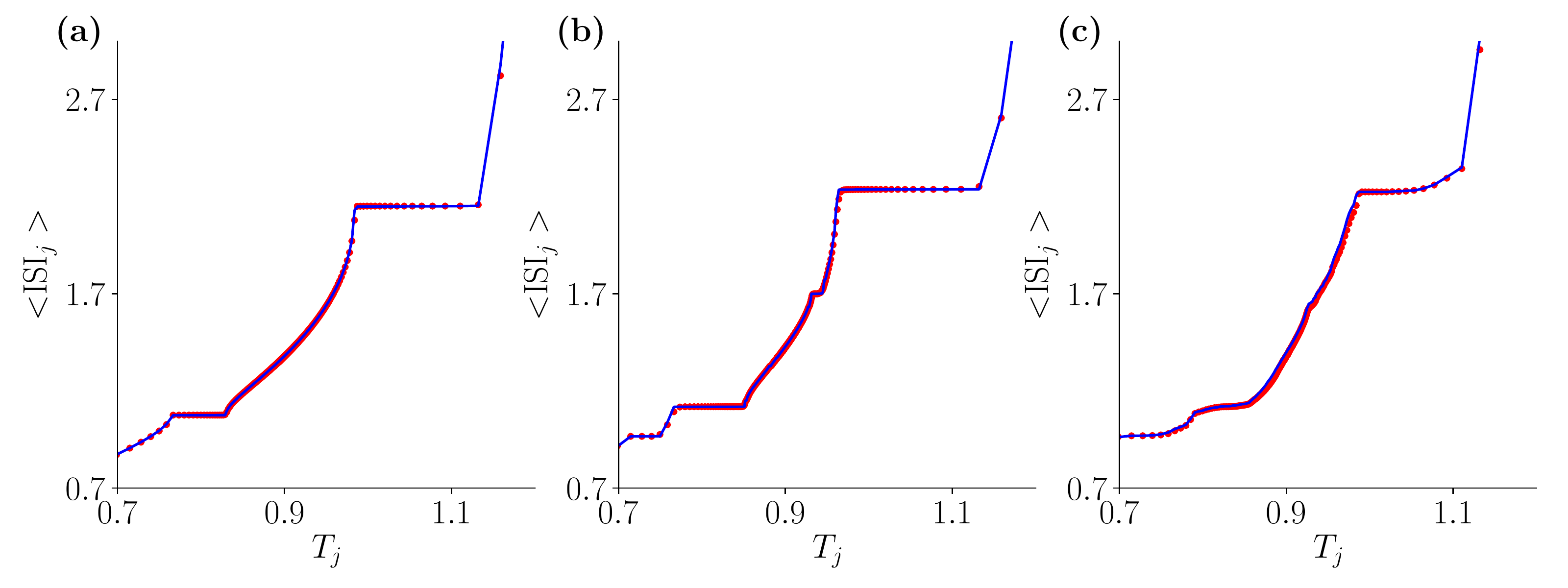}
\caption{Time-averaged coupling-modified ISIs as a function of the
intrinsic ISI for a population of 2000 QIF neurons in
three different states: (a) PS-II, (b) M-PS, and (c) collective chaos.
The red dots are
obtained with direct simulations of the population of QIF neurons, while the
blue line is obtained forcing each neuron with the FRE. Note the multiple
plateaus in the middle  panel.
Parameters are as in Fig.~\ref{Fig8}: $\sqrt{\bar{\eta}}=3.5$
and (a) $J=-9.60$; (b) $J=-10.70$; (c) $J=-11.30$.
}
\label{Fig9}
\end{figure}

To further characterize the microscopic dynamics of PS-II, M-PS and collective chaos,
in Fig.~\ref{Fig9} we calculate the 
time-averaged coupling-modified ISIs of the neurons, and plot them against 
each neuron natural ISI $T_j$. In the PS-II state shown in panel (a), 
the lower and upper plateaus correspond, respectively, to the average period
between two consecutive peaks of the mean field, and to the period of the mean
field oscillation in Fig.~\ref{Fig8}.
Here it is convenient to recall Table~\ref{table},
where the relations between macroscopic and microscopic dynamics are indicated. 

\begin{figure}[b]
\centering
\includegraphics[width=85mm]{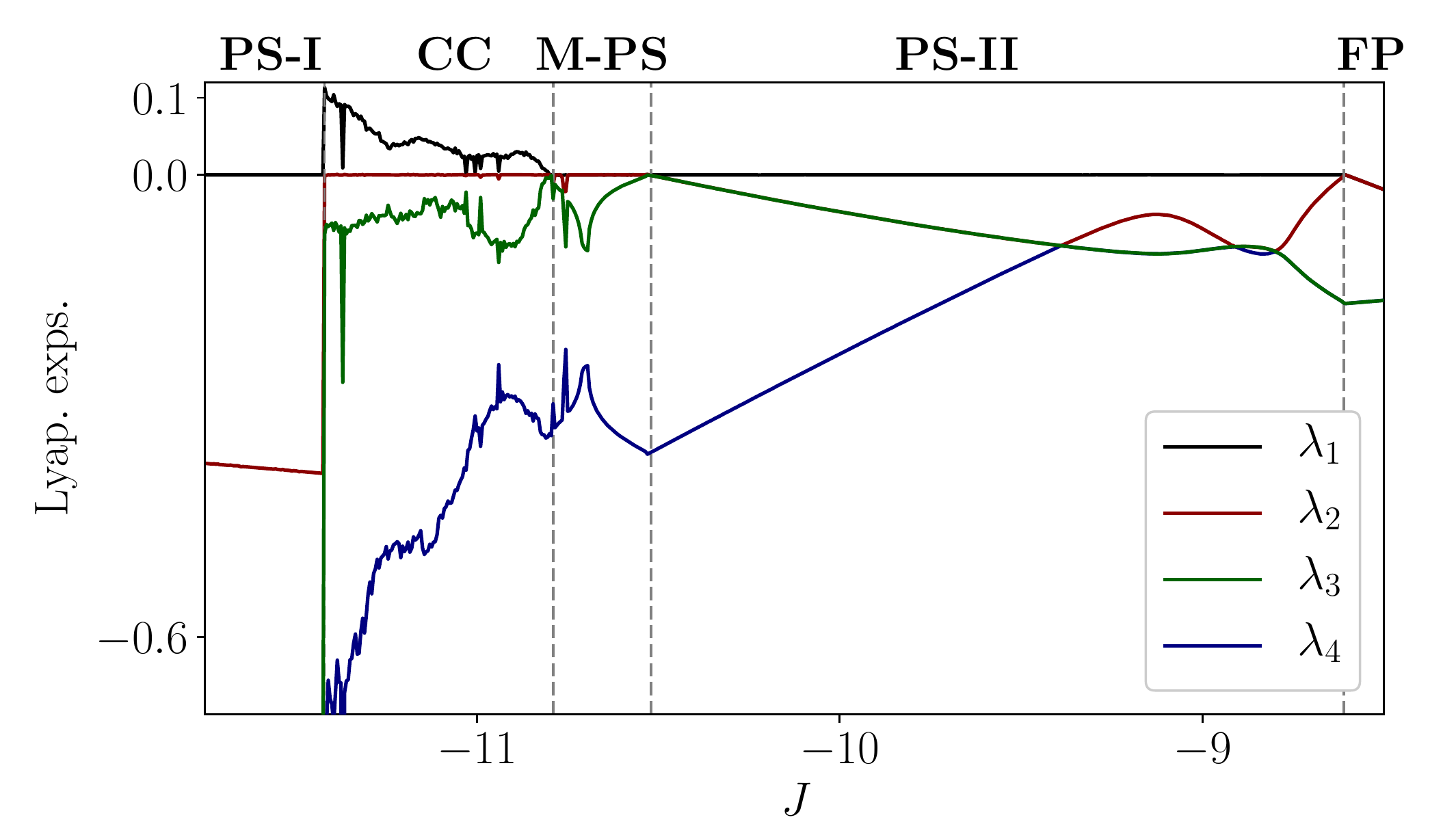}
\caption{The four largest Lyapunov exponents for $\Delta=0.1$ and
$\sqrt{\bar{\eta}}=3.5$.
The stability regions of the different attractors are indicated
by vertical gray dashed lines.}
\label{Fig10}
\end{figure}

Finally, we investigate the bifurcations that connect these partially
synchronous states, again
relying on the computation of the Lyapunov spectrum of
Eqs.~\eqref{fre1}. As we did in Section IV for identical neurons,
we evaluate the four largest Lyapunov exponents along the $J$ direction in the phase
diagrams, near the Hopf bifurcation. Figure~\ref{Fig10} reveals a scenario
qualitatively similar to the identical case (except that, at least for the specific $\bar\eta$ value adopted, no bistability was detected). 
Starting from a fixed point, the
Hopf bifurcation produces a periodic solution (PS-II) with a vanishing largest
LE, which then undergoes a Neimark-Sacker bifurcation leading to a
quasiperiodic solution (M-PS).
Eventually, this quasiperiodic solution breaks down giving rise to a chaotic state.
Finally, increasing inhibition above a critical level
makes the Lyapunov exponents to change abruptly, and chaos is suddenly
replaced by a periodic orbit (PS-I). This is in consistency with an
exterior crisis undergone by the chaotic attractor.

\subsection{Boundaries of incoherence for large heterogeneity}
At this point, we discussed a fixed value of the heterogeneity $\Delta=0.1$. 
We now study the effect of increasing values of $\Delta$ on the 
stability boundaries of incoherence. 
As previously discussed, the Hopf bifurcations become increasingly
supercritical as the level of heterogeneity grows,
and this is particularly pronounced for inhibitory coupling.
Hence the Hopf boundaries are a good proxy to
bound the regions with oscillations of either type
(PS-I, PS-II, M-PS, collective chaos).

\begin{figure}
\centering
\includegraphics[width=75 mm]{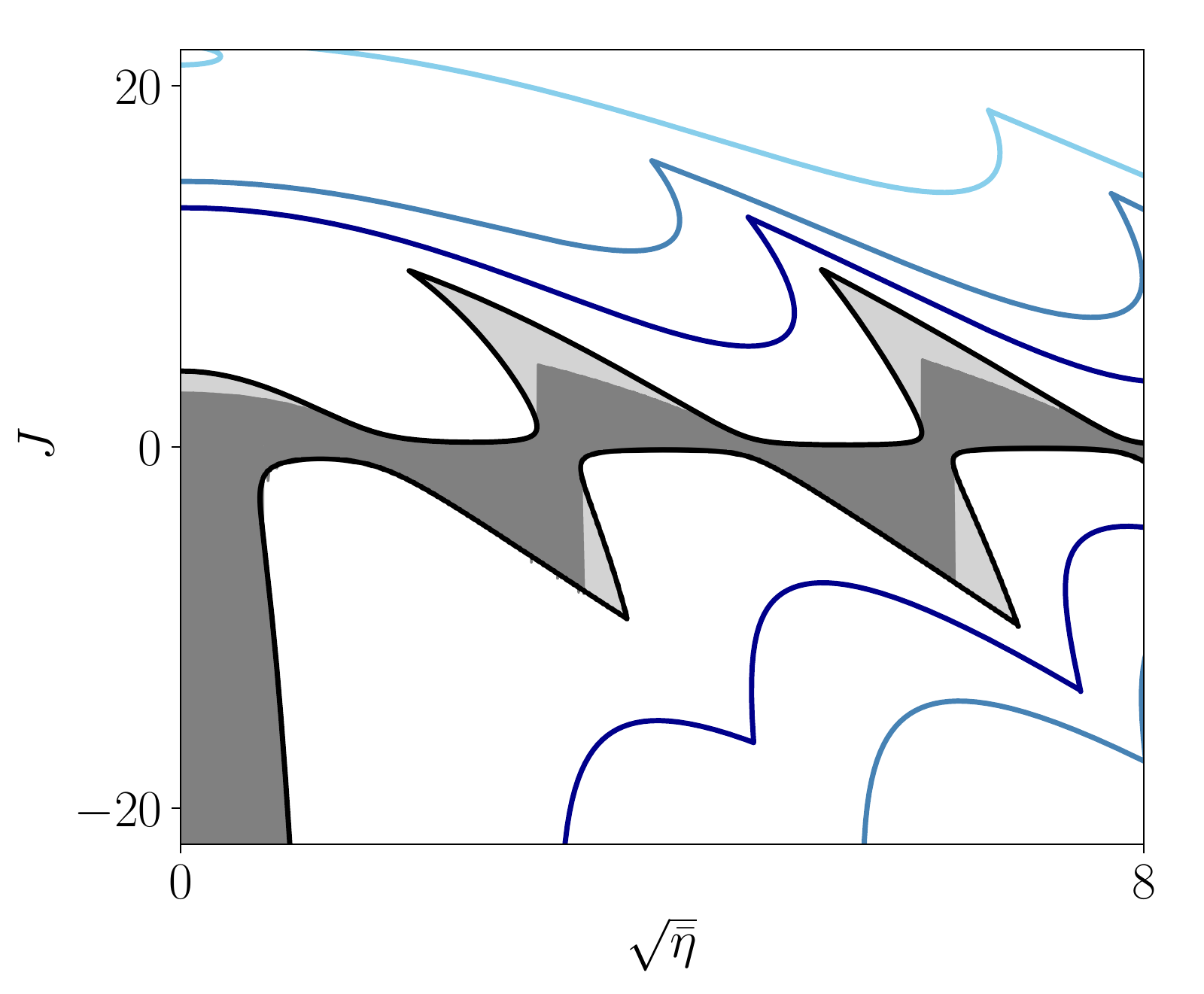}\hfill
\caption{ Increasing the level of heterogeneity $\Delta$
reveals different synchronization scenarios for excitation and inhibition (see
text). Black, dark blue, blue and light blue lines correspond, respectively,
to the Hopf boundaries of Eqs.~\eqref{fre1} with
$\Delta=0.1,~5,~10,~20$. These boundaries
determine the regions of stability of the incoherent/asynchronous states.
In the shaded regions incoherence is stable for $\Delta=0.1$. In the dark shaded
region the only attractor is incoherence.}
\label{Fig11}
\end{figure}

Figure~\ref{Fig11} shows the Hopf boundaries
increasing values of $\Delta$. Note that the region of oscillations for
inhibitory coupling progressively shrinks, and eventually disappears from the
diagram. Accordingly, given a value of
$\bar{\eta}$, there is a value of $\Delta$ for which, no matter how strong 
inhibition is, the neurons will not synchronize. The fragility of the
oscillations against heterogeneity is consistent with previous computational
studies of networks of inhibitory, conductance-based spiking neurons
\cite{WB96,WCR+98,TJ00}.
However, note that synchronization can always be achieved for strong enough
excitatory coupling. This highlights a fundamental asymmetry between the excitatory
and the inhibitory oscillatory regions in networks of QIF neurons.
We emphasize that this asymmetric behavior is not found in the
heterogeneous Kuramoto model with delay~\cite{ES03,MPS06,LOA09}.
A possible explanation for such asymmetry is that, at variance with
other self-sustained oscillators, QIF neurons
cease to oscillate for strong enough inhibition.
On the contrary, excitation just speeds up QIF neurons, 
which remain oscillatory.

\section{Conclusions and Discussion}

We analyzed the dynamics of a large population of QIF
neurons with synaptic delays. To a large extent the analysis
was carried out using the FRE Eqs.~\eqref{fre1}, which
is mathematically tractable and allows for an efficient computational analysis.
For identical neurons, we have
extended the analytical results in~\cite{PM16} to the excitable regime
($\bar \eta<0$). Our analytical predictions pointed out parameter regimes
where non-trivial dynamics should necessarily occur. In these regions of
parameters we performed an extensive numerical exploration
supported by the computation of the
Lyapunov spectrum, which revealed the existence of partially
synchronous states. One of these states, which we
called M-QPS, appears after a Neimark-Sacker bifurcation of QPS
that superimposes a second (modulating) frequency.
Partially synchronous states ---especially QPS---
coexist with full synchronization in a large region of the parameter space.
The existence in the phase diagram Fig.~\ref{Fig5}
of what looks like a second tongue for QPS
is an intriguing finding of this work.
Can its origin be understood, at least heuristically?
We finally showed that the
partially synchronized states observed in the absence of disorder
also have their counterpart in the presence of heterogeneity.
However, disorder induces diversity in the microscopic behaviors of the single
neurons.

\begin{figure}[t]
\centering
\includegraphics[width=70mm]{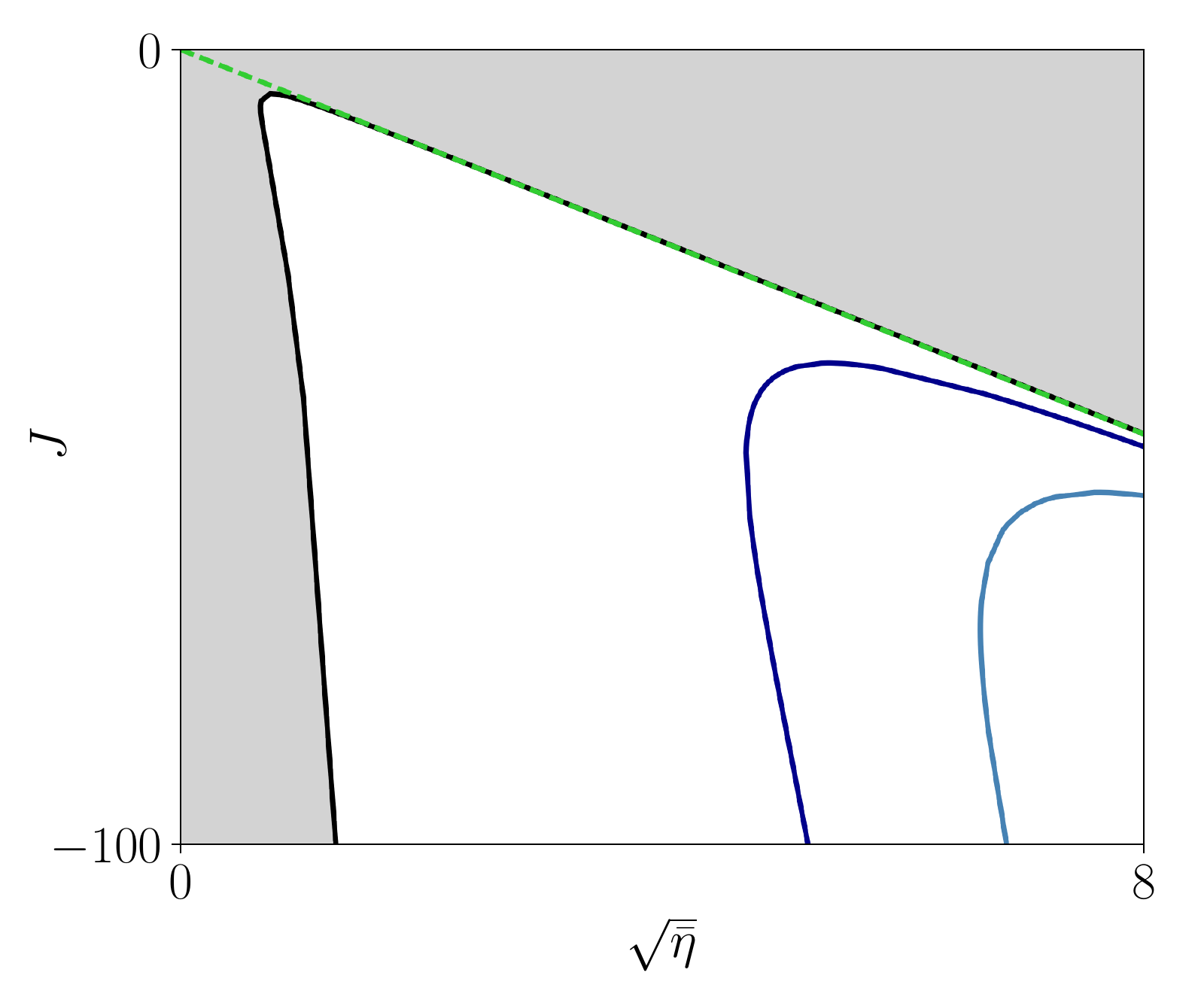}
\caption{Oscillations emerge only for inhibitory coupling in
the traditional firing rate model Eq.~\eqref{freWC}. In the gray region, limited by the
black line ($\Delta=0.1$), the fixed point determined by Eq.~\eqref{fp}
is stable and looses stability via a Hopf bifurcation
---compare with Fig.~\ref{Fig11}. The
dark blue, and blue curves respectively correspond to
$\Delta=5,~\text{and }10$. The green dashed boundary corresponds to the case
$\Delta=0$ and is a straight line.
}
\label{Fig12}
\end{figure}

To conclude, we demonstrate that most of the dynamics of the FRE Eqs.~\eqref{fre1}
investigated here cannot be
reproduced using traditional firing rate models
~\cite{WC72,DA01,GK02,ET10}.
To show this we note that the fixed points of Eqs.~\eqref{fre1}
have precisely the structure of traditional firing rate models,
while the dynamics is generically different~\cite{DRM17}.
Solving the fixed point equation corresponding to Eq.\eqref{frer} for $v$,
and substituting it into the fixed point equation corresponding to Eq.~\eqref{frev},
one obtains an equation for the steady firing rate 
\begin{equation}
r_*=\Phi(J r_*+\bar \eta).
\label{fp}
\end{equation}
The function
\begin{equation}
\Phi(x)=\frac{1}{\sqrt{2}\pi} \sqrt{x+\sqrt{x^2+\Delta^2}},
\nonumber
\end{equation}
is the so-called `transfer function' of a population of QIF neurons
with Lorentzian distribution of currents~\cite{DRM17,Lai14} 
---steady state equations for arbitrary 
distributions of currents can be obtained self-consistently,
see Eq.~(C1) in \cite{MPR15}.
Clearly, the traditional first-order firing rate model with time delays
\begin{equation}
\dot r=-r+\Phi(J r_{D=1}+\bar \eta),
\label{freWC}
\end{equation}
largely investigated in the literature
has exactly the same fixed points as Eqs.~\eqref{fre1}, but different
dynamics ---see e.g.~\cite{RBH05,BBH07,BH08,RM11,LB11,KFR17,KEK18}
for studies of Eqs.~\eqref{freWC} using different transfer functions.
Indeed, the linear stabililty analysis of the fixed points of
Eq.\eqref{freWC} gives the characteristic equation
\begin{equation}
\lambda =-1 +  \Phi' J e^{-\lambda},
\nonumber
\end{equation}
where $\lambda$ is an eigenvalue, and
$\Phi'$ is the derivative of the transfer function evaluated at the fixed
point $r_*$, determined by Eq.~\eqref{fp}.
The nonstationary instabilities (obtained using the
condition of marginal stability $\lambda=i \Omega$) are
depicted in Fig.~\ref{Fig12} for different values of
the heterogeneity $\Delta$, and clearly differ from the Hopf boundaries
of the FRE~\eqref{fre1} shown in Fig~\ref{Fig11}. Specifically,
the traditional firing rate model Eq.~\eqref{freWC} only displays oscillations for
inhibitory coupling and $\bar \eta>0$, while the FRE Eqs.~\eqref{fre1}
show oscillations for both excitation and inhibition, even for
$\bar \eta<0$ ---see Figs.~\ref{Fig1},~\ref{Fig6}, and \ref{Fig11}. 
Moreover, the tent-shaped structure of the Hopf
boundaries of Eqs.~\eqref{freWC} is lost in the 
traditional firing rate model Eq.~\eqref{freWC}.

Nonetheless note that as the heterogeneity $\Delta$
is increased, the behavior of the
Hopf boundaries of Eq.~\eqref{freWC} qualitatively agrees
with that of the FRE Eqs.~\eqref{fre1}: The region of oscillations
in both models shifts to large $\bar \eta$ values,
in consonance with the well known result that
quenched heterogeneity cannot be counterbalanced
by inhibitory coupling to produce synchronization
~\cite{WB96,WCR+98,TJ00,DRM17}. Moreover, we have
shown that for large heterogeneity the Hopf boundaries
of Eqs.~\eqref{fre1} become supercritical, and this coincides
with what is generically found in traditional firing rate models
with small delays~\cite{RM11}.
In fact, though Eq.\eqref{freWC} is heuristic,
it has proven to be remarkably effective at describing the
oscillatory dynamics of networks of spiking neurons with
strong noise~\cite{RBH05,BBH07,BH08,RM11,LB11,KFR17,KEK18,SKS+18},
and is a paradigmatic mean-field model to investigate the effect
of various types of delays in neuronal networks, see e.g.
~\cite{HA06,BK08,VCM07,CL09,FF10,Tou12,WRO+12,Vel13,VF13,FT14,DGJ+15}.

Finally, we want to note the resemblance of the partially synchronized
states investigated here with the
so-called sparsely synchronized states~\cite{BH08},
in which strong inhibition and noise produce irregular spiking but
a coherent macroscopic oscillation. Remarkably,
in both states the period of the macroscopic oscillation is determined by the time delay  but differs from the ISIs of the
single cells.
However, microscopically, the neurons have a qualitatively
different behavior: in the QPS, their dynamics is purely deterministic and
quasiperiodic, while in the sparse synchrony it is stochastic and irregular.

\acknowledgments
 
We acknowledge support by the Spanish Ministry of Economy and Competitiveness under Projects No. FIS2016-74957-P , No. PSI2016-75688-P, and No. PCIN-2015-127. We also acknowledge support by the European Union’s Horizon 2020 Research and Innovation programme under the Marie Skłodowska-Curie Grant No. 642563.

\bibliographystyle{prsty}

\end{document}